\documentclass[fleqn,usenatbib]{aa}

\DeclareRobustCommand{\VAN}[3]{#2}
\let\VANthebibliography\thebibliography
\def\thebibliography{\DeclareRobustCommand{\VAN}[3]{##3}\VANthebibliography}

\usepackage{graphicx}	
\usepackage{amsmath}	
\usepackage{amssymb}	
\usepackage{epstopdf}
\usepackage{threeparttable}
\usepackage{bm}
\usepackage{soul}
\usepackage{hyperref}
\usepackage{blindtext}
\usepackage{times}
\usepackage{txfonts}

\newcommand{\eqb}{\begin{eqnarray}}
\newcommand{\eqe}{\end{eqnarray}}
\newcommand{\ergs}{erg~s$^{-1}$}

\newcommand{\oiiil}{[O\,{\footnotesize III}] $\lambda$5007}
\newcommand{\oiii}{[O\,{\footnotesize III]}}

\newcommand{\hb}{\rm H\ensuremath{\beta}}

\newcommand{\niir}{[N\,{\footnotesize II}] $\lambda$6584}
\newcommand{\niil}{[N\,{\footnotesize II}]}
\newcommand{\feii}{Fe\,{\footnotesize II}}

\begin{document}

\title{Properties of IR selected Active Galactic Nuclei}

\author{
C. G. Bornancini\inst{1,2} \and
G. A. Oio\inst{1} \and M. V. Alonso\inst{1,2} \and D. Garc\'ia Lambas\inst{1,2}
}

\institute{Instituto de Astronom\'{\i}a Te\'orica y Experimental, (IATE, CONICET-UNC), C\'ordoba, Argentina \\
\email{cbornancini@unc.edu.ar}
\and
Observatorio Astron\'omico de C\'ordoba, Universidad Nacional de C\'ordoba, Laprida 854, X5000BGR, C\'ordoba, Argentina\\
}


\date{Received XXX; accepted YYY}

\abstract
{Active galactic nuclei (AGN) of galaxies play an important role in the life and evolution of galaxies due to the impact they exert on certain
properties and the evolutionary path of galaxies. 
It is well known that infrared (IR) emission is useful for selecting galaxies with AGNs, although it has been observed that there is contamination by star-forming galaxies. }
{In this work we investigate galaxy properties hosting AGNs identified at mid and near-IR wavelengths. The sample of AGNs selected at IR wavelengths was confirmed using optical spectroscopy and X-ray photometry.
We study the near-UV, optical, near and mid-IR (MIR) properties, as well as \oiiil\ luminosity, black hole mass and morphology properties of optical and IR colour selected AGNs.}
{We selected AGN candidates using two mid-IR colour selection techniques, a power-law emission method and a combination of mid and near-IR selection techniques. We confirm the AGN selection with two line diagnostic diagrams that use the ratio \oiii/\hb\ and the emission line width $\sigma_{\rm \oiii}$ (kinematics-excitation diagram, KEx) and the host galaxy stellar mass (mass-excitation diagram, MEx), as well as X-ray photometry.}
{According to the diagnostic diagrams, the methods with the greatest success in selecting AGNs are those that use a combination of a mid and near-IR selection technique and a power-law emission.
The method that use a combination of mid and near-IR observation selects a large number of AGNs, and is
reasonably efficient in both the success rate (61\%) and total number of AGN recovered. We also find that the KEx method presents contamination of SF galaxies within the AGN selection box.
According to morphological studies based on the S\'ersic index, AGN samples have higher percentages of galaxy morphologies with bulge+disk components compared to galaxies without AGNs.}
{}


\maketitle


\section{Introduction}

Active galactic nuclei (AGN) are compact regions at the centre of massive galaxies, perhaps all of them, that emit large amounts of radiation of non-thermal origin.
The origin of this large emission of energy has been connected with the presence of a massive black hole at the centre of the galaxies \citep{zeldo, rees}. 
Since their discovery \citep{sch}, AGNs have become a fundamental part of understanding the origin and evolution of galaxies.
The identification is fundamental for the study of host galaxy properties and to inquire about the evolutionary processes that galaxies undergo throughout their lives.
It is well known the connection between various parameters related to the black hole and its host galaxy, although there are several orders of magnitude difference in their physical sizes. We can cite among them, the relationships between the black hole mass and the bulge mass \citep{magorrian98, wandel, mclure,haring04, graham15,ding}, between black hole mass and the velocity dispersion \citep{ferrarese00,Gebhardt00, merrit01,beifiori}, and even also with the host galaxy mass \citep{bandara09}.

Some studies showed that the selection of AGNs in the UV, optical and even in the X-ray surveys missed several relatively dust obscured AGNs and almost all heavily obscured Compton-thick AGN population \citep{gilli07,daddi07, treister09}.
It is well known that the selection  in the infrared (IR) is a potentially powerful way to identify a variety of AGNs, including obscured AGNs \citep[and references therein]{hickox2018}.  Also, the spectral energy distribution (SED) of AGNs in the mid-IR (MIR) is very different than that of normal galaxies and stars \citep{assef2018}.
According to the observed redshift range, the emission in the IR can come from structures close to the black hole such as the torus (observed at low redshifts, i.e., z$\leq$1.5) as well as from the accretion disk (at z$\geq$ 1.5, \citealt{chung2014, assef2018b}). 
After the advent of the $Spitzer$ $Space$ $Telescope$, large samples of AGNs could be obtained with the IRAC camera that operated with filters centred at $[3.6]$, $[4.5]$, $[5.8]$ and $[8.0]$ $\mu$m \citep{L04, S05, hickox2007, donley07, donley08, park2010, assef2011, donley12, mendez2013, bornan2017, Ch17, bornan2018, bornan2020}. Within the first investigations, we can cite \citet{Eisen} who discovered a sequence of objects in the $[3.6]-[4.5]$ vs. $[5.8]-[8.0]$ colour-colour diagram
 formed by compact objects identified in the [3.6] $\mu$m filter. 
 Preliminary spectroscopy of these unresolved sources suggested that they are a mixture of broad-lined quasars (QSOs) and starburst galaxies \citep{Eisen}.
 Following this idea, \citet{L04} identified the position of QSOs detected in the Sloan Digital Sky Survey (SDSS) in a log(S$_{8.0}$/S$_{4.5}$) vs. log($S_{5.8}$/S$_{3.6}$) (where  S$_{\nu}$ is the flux density at the frequency $\nu$) diagram and presented a colour-cut criterion in order to select them.
 Another well-known MIR based method of AGN selection was that presented by \citet{S05}. 
 These authors used a similar approach to establish an empirical criterion to separate active galaxies from other sources based on the distribution of nearly 10,000 spectroscopically identified sources from the AGN and Galaxy Evolution Survey \citep{AGES}.
 It is well known that in the MIR spectra, 
 AGNs are often characterised  by a power-law (S$_{\nu} \propto \nu^{\alpha}$, where $\alpha$ is the spectral index) in their SEDs at rest-frame optical, near and MIR wavelengths \citep{neugebauer1979, elvis,alonso-herrero06, donley07, Ch17}. 
This emission can be originated by non-thermal processes in the nuclear region and by thermal emission due to various nuclear dust components \citep{rieke81}.

The optical and UV light emitted by the accretion disk located near the black hole is absorbed by a dust structure around them, which reprocesses it and re-emits the radiation at IR wavelengths. 
According to the Unified Model \citep{anto93,urry95}, this structure formed by dust would have the appearance of a torus, although several studies suggested that the shape is not so regular \citep{alonso-herrero11, audibert17} and it could even have considerable sizes \citep{gould12, donley18}. The re-emission of the radiation produced can be described with a power law from 1 to 10 $\mu$m (which is the coverage sampled by the 4 IRAC filters), showing a thermal bump which peaks around 10 $\mu$m. Some authors stated that AGNs have a variety of slopes ranging from $\alpha$ $=-$1 for selected QSOs in the optical \citep{elvis}, while others suggested that QSOs have indices ranging from $-$0.5 to $-$2 \citep{ivezic2002, barmby06}. Whereas for star-forming (SF) galaxies, they were expected to have $\alpha$ $\simeq +2$ \citep{barmby06}. This method was used by several authors to select samples of AGNs \citep{L04, alonso-herrero06, polletta2006, L07, donley07, donley08, park2010, donley12, Ch17, bornan2020}.

It is well-known that some IR selection methods for AGNs also select other galaxy types as contaminants as well as failing to detect some AGN types.
\citet{mendez2013} studied the $Spitzer$/IRAC and X-ray selection methods for AGNs identified in four large fields such as of the Cosmological Evolution Survey (COSMOS, \citealt{scoville}), the XMM Large Scale Structure survey (XMM-LSS, \citealt{pierre2004}), the European Large Area $Infrared$ $Space$ $Observatory$ (ES1-S1, \citealt{oliver00}) and Chandra Deep Field South (CDFS, \citealt{giacconi2001}) surveys. 
They found that the selected sample of galaxies according the \citet{S05} method is contaminated by star-forming galaxies at z$\sim$0.3 and by quiescent galaxies  at z$\sim$1.1. 
As noted by \citet{donley12}, the two selection methods proposed by \citet{L07} and \citet{S05} suffer from contamination by galaxies classified as pure starburst determined by $Spitzer$ InfraRed Spectrograph observations. Although the effect is greater at high redshifts, it is also observed for galaxies in the redshift range of 
0.5 $< z <$ 1.
Based on simulations, \citet{sajina2005} found a larger amount of contamination from intermediate-redshift polycyclic aromatic hydrocarbon dominated sources near the limits of the \citet{L04} AGN selection region.   They are commonly identified with a wide variety of objects from dusty starbursts, quiescent and low metallicity galaxies. 
\citet{geor2008} found that AGN selection methods based on either colours or power-law spectra would fail to detect a large fraction of Compton-thick AGN candidates.
In a similar way, \citet{park2010} analysing the properties of power-law and X-ray emission from AGNs found that only 22\% of the X-ray AGNs are detected by the power-law AGN selection method.

{In this paper we analyse the properties of galaxies hosting AGNs identified at IR wavelengths and checked by means of optical spectroscopy and X-ray photometry.
For this we will use four MIR and near-IR methods according to the criteria of \citet{L04, S05, Ch17} and \citet{M12} and two line diagnostic diagrams that use the quotient \oiii/\hb\ and the emission line width $\sigma_{\rm \oiii}$ \citep{Zh18} and the host galaxy stellar mass \citep{J14}}.

This paper is organised as follows: 
 In Section~\ref{data}, we present all datasets used in this study and in Section~\ref{selection} we detail the different MIR and near-IR ($K_s$) AGN selection methods. In Section~\ref{kex}, we analyse the KEx and MEx line diagnostic diagrams for AGNs. In Section~\ref{nearuv} we explore the X-ray emission and the different MIR and near-IR methods and line diagnostic diagrams properties for AGNs. We study the properties of AGNs selected by KEx and MEx diagnostic diagrams in Section 6.  
 In Section 7 we study the X-ray properties, black hole mass, accreting properties and the morphology of AGN selected through the MEx diagram. And finally, the summary and discussion of our study are presented in Section~\ref{sum}.

Throughout this work we will use the AB magnitude system \citep{oke} and we will assume a $\Lambda$CDM cosmology with H$_{0} = 70$ km s$^{-1}$ Mpc$^{-1}$, $\Omega_{\rm M} = 0.3$, $\Omega_{\rm \Lambda} = 0.7$.

\section{Datasets}

\label{data}

Our main goal is to study the galaxy properties of four different MIR and near-IR selection methods and two line diagnostic diagrams that use the emission line quotient \oiii/\hb\ and the emission line width $\sigma_{\rm \oiii}$ (KEx diagram, \citealt{Zh18}) and  \oiii/\hb\ vs. stellar mass (MEx diagram, \citealt{J11,J14}). 
For this study, it is necessary to  consider catalogues with spectral data in order to measure line widths and emission line quotients. We selected our AGN sample in the COSMOS field \citep{scoville} from the COSMOS2015\footnote{The catalogue can be downloaded from \url{ftp://ftp.iap.fr/pub/from\_users/hjmcc/COSMOS2015/}} catalogue \citep{laigle} and from the zCOSMOS redshift survey \citep{lilly07,lilly09}. 
The COSMOS is a deep, wide area, multi-wavelength survey that contains observational information of more than 1 million galaxies over a 2 square degrees region centred at (RA,Dec) = (10$^h$ 00$^m$ 28.6$^s$, 2$^\circ$ 12' 21''). 
The COSMOS2015 \citep{laigle} catalogue provides photometric data from 0.24$\mu$m to 500$\mu$m, that includes GALEX FUV and in the near-UV ($NUV$) observations \citep{zamo, capak07}; $u^*$, $g$, $r$, $i$, $z$, $J$, $H$, $K_s$ taken with the CFHT/MegaCam \citep{sanders07}; $B$, $V$, $g$, $r'$, $i^+$, $z^{++}$ broad bands; IA427, IA464, IA484, IA505, IA527, IA574, IA624, IA679, IA709, IA738, IA767, IA827, intermediate bands and NB711, NB816 narrow bands from the Subaru Suprime-Cam \citep{tani07,tani15}; near-IR $Y$, $J$, $H$, $K_s-$band data taken with WIRCam and Ultra-VISTA \citep{mccrac10,mccrac12}; six $Spitzer$ IRAC/MIPS bands: 3.6, 4.5, 5.8, 8.0, 24 and 70 $\mu$m \citep{sanders07,lefloch09,ashby13,ashby15} and Herschel PACS/SPIRE 100, 160, 250, 350, and 500 $\mu$m \citep{oliver12,lutz11}. 

zCOSMOS \citep{lilly09} is a large spectroscopic survey obtained through more than 600 hours of observations with the Visible Multi Object Spectrograph (VIMOS) mounted on the Very Large Telescope (VLT) at the European Southern Observatory (ESO) in Chile. This redshift survey is divided in two parts: the zCOSMOS-$bright$ and the zCOSMOS-$deep$. 
The first was designed to yield a high and fairly uniform sampling rate (about 70\%, \citealt{knobel12}), with a high success rate in measuring redshifts approaching 100\% at 0.5$ < z < $0.8, covering the	approximately 1.7 deg$^2$ of the COSMOS field. The second part, zCOSMOS-$deep$, observed a	small number of galaxies selected to mostly lie at higher	redshifts,	1.5$ < z < $3.0.	
In this work we used the last spectroscopic release zCOSMOS (DR3) that contains redshift data from 20689 galaxies at 0.2 $< z <$ 1.2 selected to have I$_{AB} <$ 22.5 mag. This catalogue provides a full set of extracted 1-dimensional spectra, plus a catalogue which contains the 1-D spectra filenames, the I-band magnitudes used for the selection, as well as the measured redshifts and the redshift confidence class parameter.	

There are several spectroscopic redshift catalogues such as DEIMOS 10k \citep{hasinger18}, MUSE Wide survey \citep{urrutia19}, FMOS-COSMOS \citep{silverman15,kashino19} and the MOSFIRE Deep Evolution Field Survey  \citep{kriek15} (see \citealt{alarcon20} for a complete sample list). We have only used the zCOSMOS-$bright$ survey because it represents a homogeneous catalogue using the same selection criteria and with a very good completeness. 

We have first selected sources within the spectroscopic redshift range $0.3\leq z_{\rm sp}\leq 0.9$ from the zCOSMOS DR3-$bright$ catalogue with good redshift estimates. 
This catalogue provides a confident class parameter which ranges from insecure and probable redshift (Class 1 and 2, respectively), one broad AGN redshift (Class 18), one line redshift (Class 9) and secure and very secure redshift (Class 3 and 4, respectively).
Each confident class is also assigned a confidence decimal, which is derived from repeat observations and by the consistency or otherwise with photometric redshifts. The confidence decimal ranges from .1 (spectroscopic and photometric redshifts are not consistent at the level of $0.04(1+z)$), .3 (special case for Class 18 and 9, consistent with photo-z only after the	redshift is changed to the alternate redshift), .4  (no photometric redshift available) to .5, (spectroscopic redshift consistent within $0.04(1+z)$ of the photometric redshift). 
In this work we have selected all sources with 3.x and 4.x, where x can take the values 1, 3, 4 and 5.
The lower limit of the redshift range used in this work, which is $z = 0.3$ corresponds to the limit used by the KEx and MEx line diagnostic diagrams proposed by \citet{Zh18} and \citet{J14}, while the upper limit of $z = 0.9$ was chosen because it is the detection limit of the observed [OIII]$\lambda$5007 line in the spectral range 5550-9450 \AA~of the VIMOS spectrograph.
We call this sample of galaxies with good spectroscopic redshift determinations as the zCOSMOS-$good$, which contains 8633 sources. 
We then correlated these spectroscopic sources with those of the zCOSMOS15 \citep{laigle} catalogue in order to obtain the stellar mass (estimated using the MASS\_BEST parameter) and other photometric data, such as rest-frame absolute magnitudes $M_{NUV}$, $M_r$, $M_J$, $M_{K_{s}}$ and the four IRAC bands.  

\section{MIR and near-IR AGN sample selection}
\label{selection}
In order to study the selection of AGNs at different MIR and near-IR wavelengths, we used four methods from \citet{L04,S05,Ch17} and \citet{M12}. The first three MIR methods use IRAC bands ($[3.6]$, $[4.5]$, $[5.8]$ and $[8.0]$ $\mu$m) and we refer throughout the paper as: L04, S05 and Ch17, while the method of \citet{M12} (M12) use a combination of $K_s+$IRAC colour criterion (KI criterion).  
L04 studied the colour distribution of objects according to the flux ratios using the 4 IRAC bands of the $Spitzer$ $Space$ $Telescope$: the ratio 8.0/4.5 $\mu$m vs the ratio 5.8/3.6 $\mu$m. In this diagram they identified those objects that have a blue continua from those with a red continua. They observed two clear sequences: one formed by objects with blue colours in $S_{5.8}/S_{3.6}$ and very red colours in $S_{8.0}/S_{4.5}$. The first sequence was identified with galaxies at low redshifts ($z < 0.2$) and the second one had red colours in both pairs of filters and its location matched the region occupied by QSOs identified in the SDSS survey.

\begin{figure*}
  \includegraphics[width=200mm]{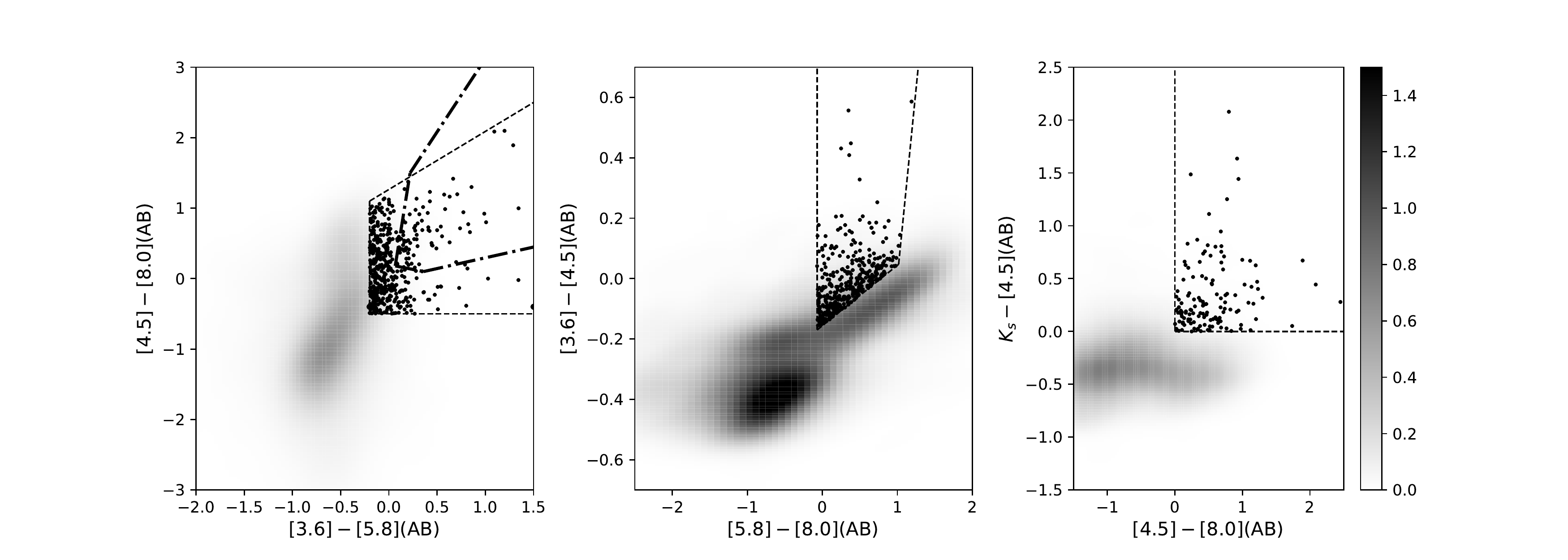}
\caption{MIR and near-IR colour-colour diagrams and the selection of AGNs for different photometric methods. Left panel shows the criterion of \citet{L04} (dotted lines) and \citet{Ch17} (dashed-point lines). Middle panel shows the criterion of \citet{S05} and right panel, the criterion of \citet{M12}. 
In each figure we show the 2D density maps that represent galaxies with spectroscopic redshifts with confidence class 3 and 4 taken from the zCOSMOS DR3 survey. The pre-selected AGNs are represented with black circles.  
}
\label{fig1}
\end{figure*}

In Figure~\ref{fig1}, we plot the 2D density maps representing MIR and near-IR colours for galaxies in the zCOSMOS-$good$ sample.
In the left panel we plot the $[4.5]-[8.0]$ vs $[3.6]-[5.8]$ colour-colour diagram. The dotted line box represents the area proposed by \citet{L04} used to select AGN samples according to the following relation:

\begin{eqnarray}
\rm log(S_{5.8}/S_{3.6})>0.1, \nonumber\\
\rm log(S_{8.0}/S_{4.5}) > -0.2 \wedge \nonumber\\
\rm log(S_{8.0}/S_{4.5}) \leq  0.8\times \rm log(S_{5.8}/S_{3.6})+0.5
\end{eqnarray}

where $\wedge$ is the logical AND operator.


In the same panel of the figure, we also plot the \citet{Ch17} colour-colour cuts of AGNs selected by their MIR power-law emission (dashed-point line region). They defined power-law sources whose IRAC four-band SEDs is well fit by a line of slope $\alpha < -0.5$, where S$_{\nu} \propto$ $\nu^{\alpha}$. These authors defined a selection box that groups the vast majority of AGNs with a power-law emission.
We used the Fig. 1 of \citet{Ch17} to obtain limiting lines 
to define the AGN selection box, which has the following form:  

\begin{eqnarray}
\Bigl([4.5]-[8.0]\Bigr)< 2.22\times\Bigl([3.6]-[5.8]\Bigr)+1.01, \nonumber\\
\Bigl([4.5]-[8.0] < 8.67\times\Bigl([3.6]-[5.8]\Bigr)-0.38, \nonumber\\ \Bigl([4.5]-[8.0]>-0.27\times \Bigl([3.6]-[5.8]\Bigr)+0.2, \nonumber\\
\Bigl([4.5]-[8.0]\Bigr) > 0.31\times \Bigl([3.6]-[5.8]\Bigr)-0.06,
\end{eqnarray}

In the middle panel of Figure~\ref{fig1}, we show the colour-colour $[3.6]-[4.5] $ vs. $[5.8]-[8.0]$ diagram  with the selection box represented by dashed lines showing the selection criterion of \citet{S05}. These authors proposed this criterion based on previous results found by \citet{Eisen} who noticed a vertical spur in the diagram formed mostly by sources spatially unresolved in the 3.6 $\mu$m images and the location of broad and narrow band AGNs selected from the AGES survey. 
These authors proposed the following empirical criteria to separate AGNs from other sources identified in the AEGIS survey\footnote{Magnitudes presented in \citet{S05} are referred to the Vega system and they were converted to the AB system using the relations taken from 
\citet{zhu}: ($[3.6]$, $[4.5]$, $[5.8]$, $[8.0]$)$_{\rm AB}$ = ($[3.6]$, $[4.5]$, $[5.8]$, $[8.0]$)$_{\rm Vega}$ + (2.79, 3.26, 3.73, 4.40).}:

\begin{eqnarray}
 \Bigl([5.8]-[8.0] \Bigr)>-0.07 \wedge\nonumber\\
\Bigl([3.6]-[4.5]\Bigr) >
0.2\times\Bigl([5.8]-[8.0]\Bigr)-0.15\wedge \nonumber\\
\Bigl([3.6]-[4.5]\Bigr)>
2.5\times\Bigl([5.8]-[8.0]\Bigr)-2.3.
\end{eqnarray}

And finally, the right panel shows the combined  near-IR and MIR $K_s-[4.5]$ vs. $[4.5]-[8.0]$ galaxy colours and the selection criteria according to \citet{M12}. In the figure, the pre-selected AGNs are represented by black circles. 
These authors found that this criterion is ideal as AGN/non-AGN diagnostics at $z\lesssim1$ based on the predictions by state-of-the-art galaxy and AGN templates.
The selection criterion is defined by the following simple conditions:

\begin{eqnarray}
 \Bigl(K_s-\left[4.5\right]\Bigr)>0 \wedge \Bigl(\left[4.5\right]-\left[8.0\right]\Bigr)>0
 \end{eqnarray}

According to the aforementioned methods, we have pre-selected the following number of candidates for AGNs: 490, 78, 362 and 133 sources conforming to the criteria of L04, Ch17, S05 and M12, respectively (see Table \ref{table0}).\\

\begin{table}
\center
\caption{Number of AGN candidates according to the MIR and near-IR methos of L04, Ch17, S05 and M12.}
\begin{tabular}{lcccc}
\hline
IR method & L04 &  Ch17 & S05 & M12\\
\hline
AGN candidates     & 490 & 78   &362  &133   \\
\hline
\label{table0}
\end{tabular}
\end{table}

We analyse the completeness of the spectroscopic sample as a function of the IR magnitudes and colours in the Appendix.

\citet{donley12} presented a different AGN selection criterion according to a power-law emission observed in the IRAC MIR bands. This method has not been used in our work since it only selects a small percentage of AGNs.
The area determined by this criterion is smaller and it is contained within the selection criteria of Ch17, pre-selecting only 36 AGN candidates. 


\begin{figure*}
\includegraphics[width=85mm]{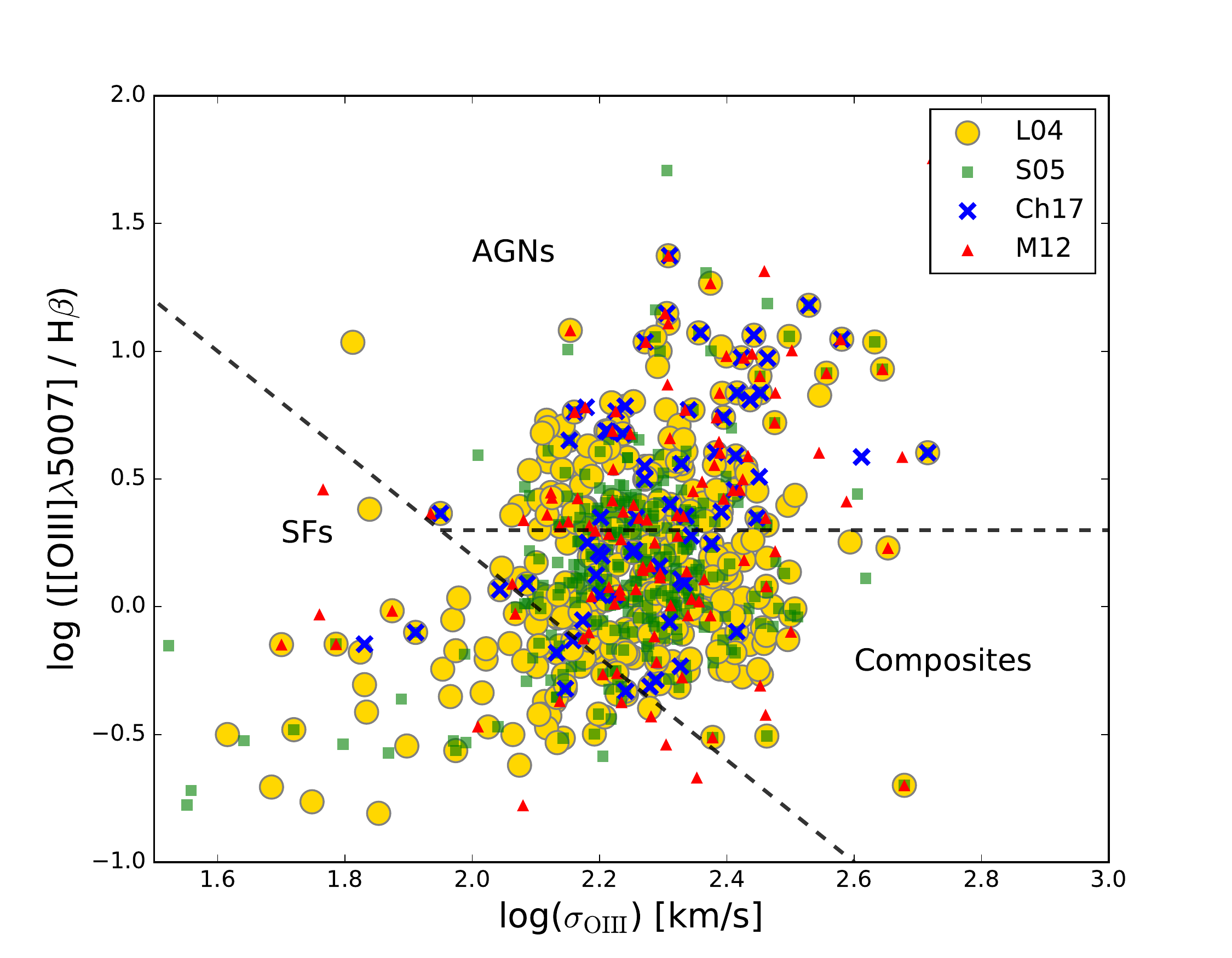}
\includegraphics[width=85mm]{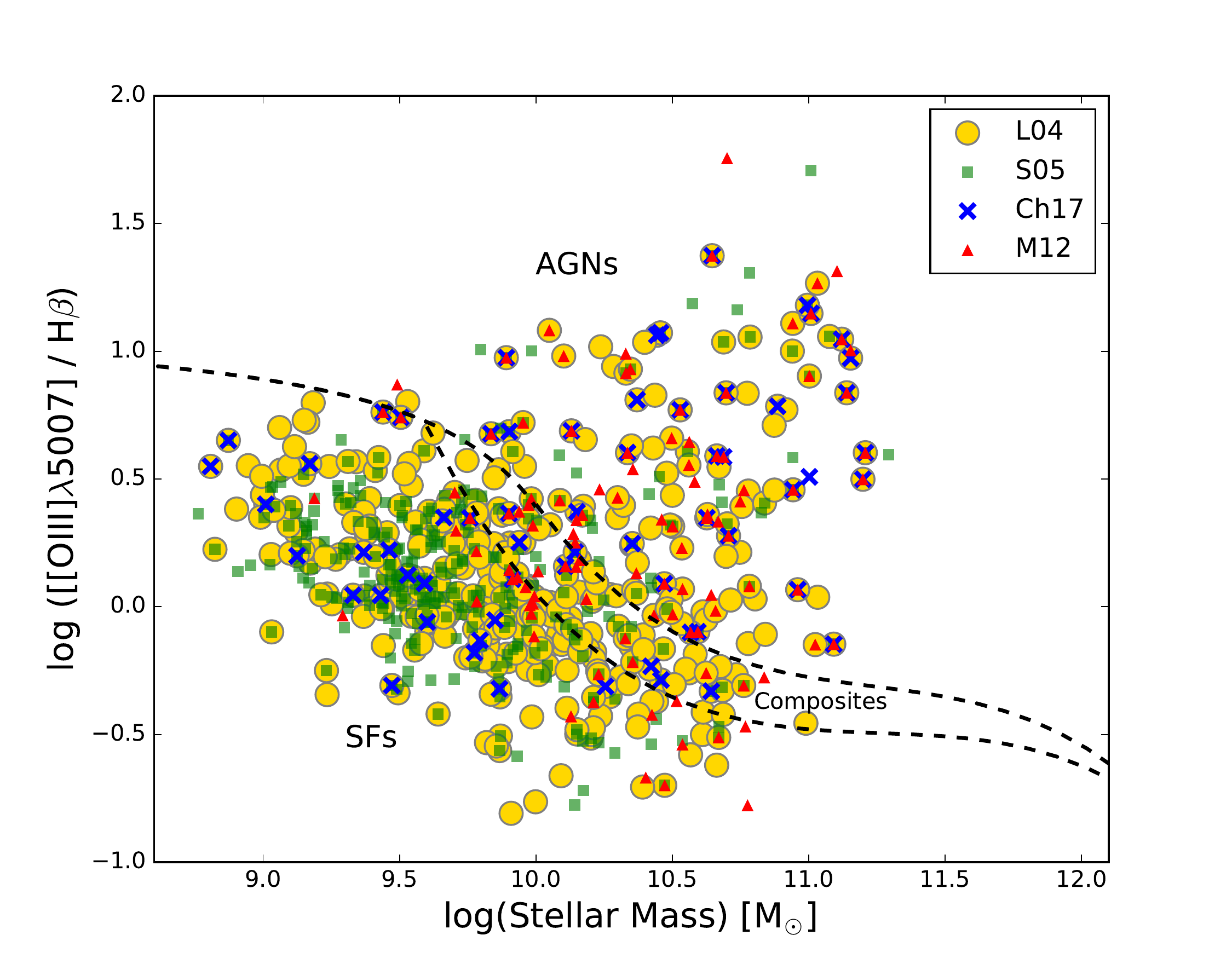}
\caption{Emission-line diagnostic diagrams. The left panel shows the kinematics-excitation (KEx) diagram and the right panel, the mass-excitation (MEx) diagram. The circles, squares, crosses and triangles represent pre-selected AGNs according to the methods of L04, S05, Ch17 and M12, respectively.  
The regions in both diagrams marked with dashed lines show the location of AGNs, composites and star-forming galaxies.}
\label{fig2}
\end{figure*}

\section{The KEx and MEx diagnostic diagrams}
\label{kex}

\citet{J11,J14} and \citet{Zh18} presented the two line diagnostic diagrams to classify and separate AGNs from star-forming and composite galaxies.
First, \citet{J11} introduced the MEx diagnostic diagram to identify AGNs in galaxy samples at intermediate redshift. These authors used a modified version of the Baldwin-Phillips-Terlevich (BPT, \citealt{bpt}) diagram, which involves \niir/H$\alpha$ and \oiiil/\hb\ quotient lines  to separate AGNs, star-forming and composite galaxies.
Since the quotient \niir/H$\alpha$ moves to the near-IR wavelengths for $z > 0.4$, they proposed to replace it for the host galaxy stellar mass.
\citet{J14} introduced a small correction to the 
line ratios between AGNs and other galaxy types. 
This new calibration was obtained using the SDSS-DR7 instead of older DR4 release \citep{J11} and an emission line signal-to-noise criterion that is applied to the line ratios rather than individual lines. In this work we will use the criterion of \citet{J14}.

In a similar way, \citet{Zh18} proposed the KEx diagram to diagnose the ionisation source and physical properties of AGNs and star-forming galaxies.  This approach uses the \oiiil/\hb\ line ratio and the \oiiil\ emission line width ($\sigma_{\rm \oiii}$) instead of stellar mass as \citet{J11,J14}.  This approach shares similar logic of the \citet{J11,J14} work, which they proposed to replace \niil/H$\alpha$ in the BPT diagram with the $\sigma_{\rm \oiii}$ to separate AGNs from star-formation using the main galaxy sample of SDSS DR7 to calibrate the diagram at low and high redshifts. 

In order to select AGNs, we used custom fitting tasks adapted from the {\small IRAF}\footnote{IRAF (Image Reduction and Analysis Facility, \citealt{tody1993}) is distributed by the National Optical Astronomy Observatories, which is operated by the Association of Universities for Research in Astronomy (AURA) under cooperative agreement with the National Science Foundation
(NSF).} $splot$ package to measure the \oiiil\ and \hb\ emission lines of the AGN candidates obtained from the pre-selection
methods in the MIR and near-IR wavelengths. For these spectroscopic measurements, we have only considered those with signal-to-noise  greater than 3 in both \oiiil\ and \hb\ lines.
All lines were well fitted using Gaussian profiles.
After applying this criteria we have observed that 18, 8, 18 and 23\% 
of the spectral line measurements were rejected using the methods proposed by L04, S05, Ch17 and M12, respectively. 
Some objects were rejected for having low S/N values while others for observing the presence of only one of the \oiiil\ or \hb\ emission lines.
The final number of AGN candidates with measured lines is 400, 64, 332 and 102 according to the methods of L04, Ch17, S05 and M12.

In Figure~\ref{fig2}, we plot the KEx (left panel) and the MEx (right panel) diagrams for AGN candidates pre-selected using the L04, S05, Ch17 and M12 methods. We have also demarcated in each panel the areas that separate the star-forming galaxies from the composite and AGN samples, according to \citet{Zh18} and \citet{J14}, respectively.
For the KEx diagram, the empirical lines are:

\begin{eqnarray}
\log(\rm \oiii/\hb) =-2\times \log \sigma_{\oiii} +4.2,\nonumber\\
\rm \wedge \nonumber\\
\log(\rm \oiii/\hb) =0.3.
\end{eqnarray}

The demarcation lines used in the MEx diagram \citep{J14} are the following:

\begin{equation}\label{eq:up}
y = \left\{ \begin{array}{ll}
      0.375/(x - 10.5) + 1.14 & \mbox{if x\ensuremath{\leq}10} \\
      a_0 + a_1 x + a_2 x^2 + a_3 x^3 & \mbox{otherwise,}
      \end{array}
     \right. 
\end{equation}

where $y \equiv$ log(\oiiil/\hb) and $x \equiv$ log$(M_{\star})$ and
the coefficients are $$\{a_0, a_1, a_2, a_3\} = 
\{410.24, -109.333, 9.71731, -0.288244\}$$
Similarly, the lower curve is given by:
\begin{equation}\label{eq:lo}
y = \left\{ \begin{array}{ll}
      0.375/(x - 10.5) + 1.14 & \mbox{if x\ensuremath{\leq}9.6} \\
      a_0 + a_1 x + a_2 x^2 + a_3 x^3 & \mbox{otherwise.}
      \end{array}
     \right. 
\end{equation}
The coefficients are $$\{a_0, a_1, a_2, a_3\} = \{352.066, -93.8249, 8.32651, -0.246416\}$$

We quoted in Tables~\ref{table1} and~\ref{table2} the percentage of AGNs, composite and star-forming galaxies according to the methods proposed by L04, Ch17, S05 and M12, using the KEx and MEx diagrams, respectively. The final numbers of pure-AGNs for the different selecting criteria of L04, Ch17, S05 and M12 are: 132, 36, 113 and 54 objects using the KEx diagram and 101, 33, 67, 63 objects using the MEx diagram.
\begin{table}
\center
\caption {Percentage of AGNs, composite and star-forming galaxies selected using MIR and near-IR methods according to the KEx criteria \citep{Zh18}.}

\begin{tabular}{lcccc}
\hline
Type & L04 &  Ch17 & S05 & M12\\
\hline
AGNs      &33\% & 56\% & 34\% & 53\%  \\
Composites &49\% & 30\% & 52\% & 34\% \\
SFs        &18\% & 14\% & 14\% & 13\% \\
\hline
\label{table1}
\end{tabular}
\end{table}

\begin{table}
\center
\caption{
Percentage of AGNs, composite and star-forming galaxies selected using MIR and near-IR methods according to the MEx criteria \citep{J14}.}
\begin{tabular}{lcccc}
\hline
Type & L04 &  Ch17 & S05 & M12\\
\hline
AGNs      &25\% &  51\% & 20\% & 61\%  \\
Composites & 15\% & 10\% & 9\% & 14\% \\
SFs        &60\% &  39\% &71\% &  25 \% \\
\hline
\label{table2}
\end{tabular}
\end{table}

\begin{figure}
  \includegraphics[width=100mm]{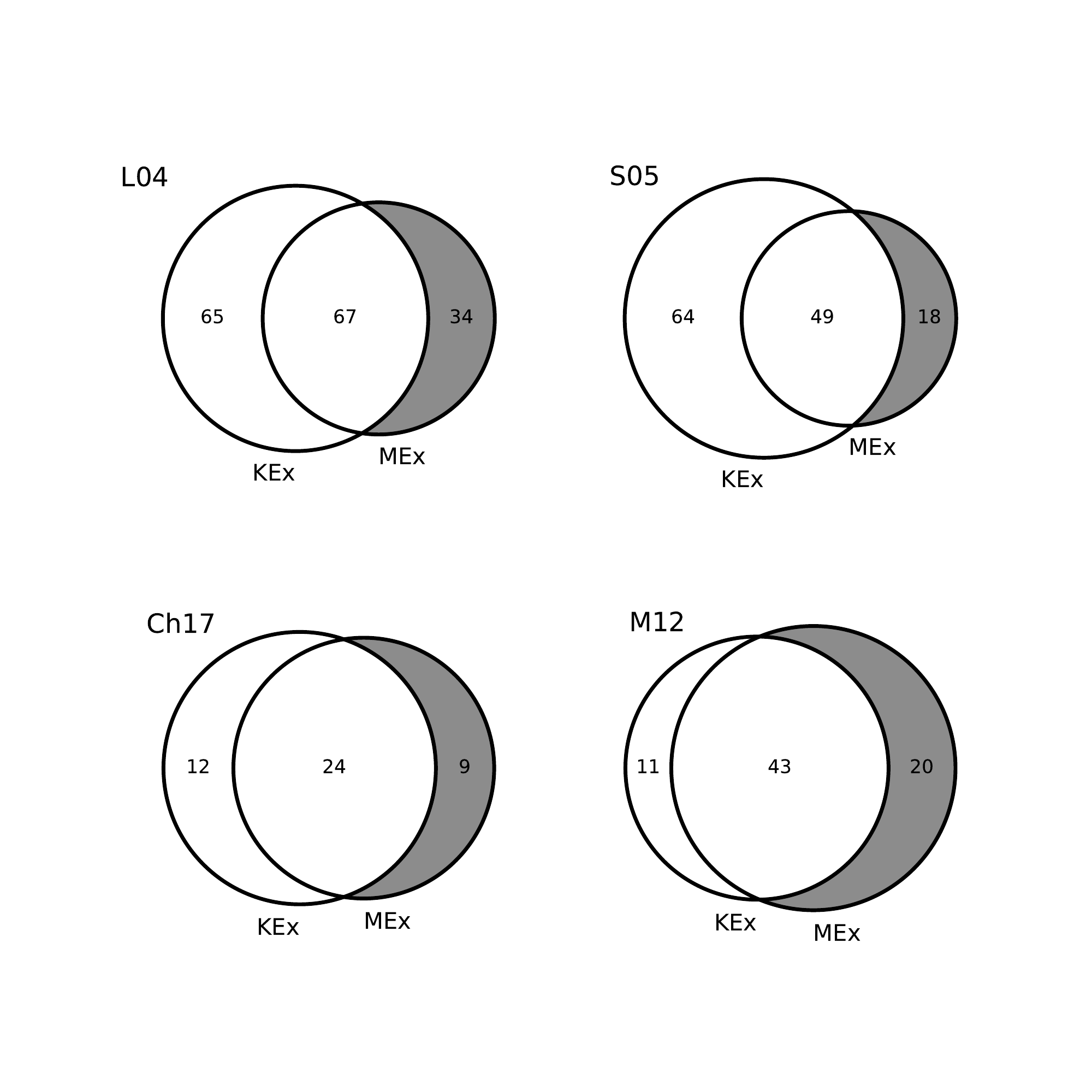}
	\caption{Venn diagrams showing the number and overlap for pre-selected AGNs according to the L04, S05, Ch17 and M12 methods and also selected using the KEx and MEx diagrams. The area and overlap of each circle are proportional to the total number of each sample. Areas that represent AGNs only identified according to the MEx diagram are shown in gray.}
\label{venn}
\end{figure}

In Figure \ref{venn} it can be seen the Venn diagrams for the selected AGNs using the KEx and MEx diagrams, according to the different selection methods in MIR and near-IR.  The methods with the highest overlap are those of Ch17 and M12.

\begin{figure*}
\includegraphics[width=80mm]{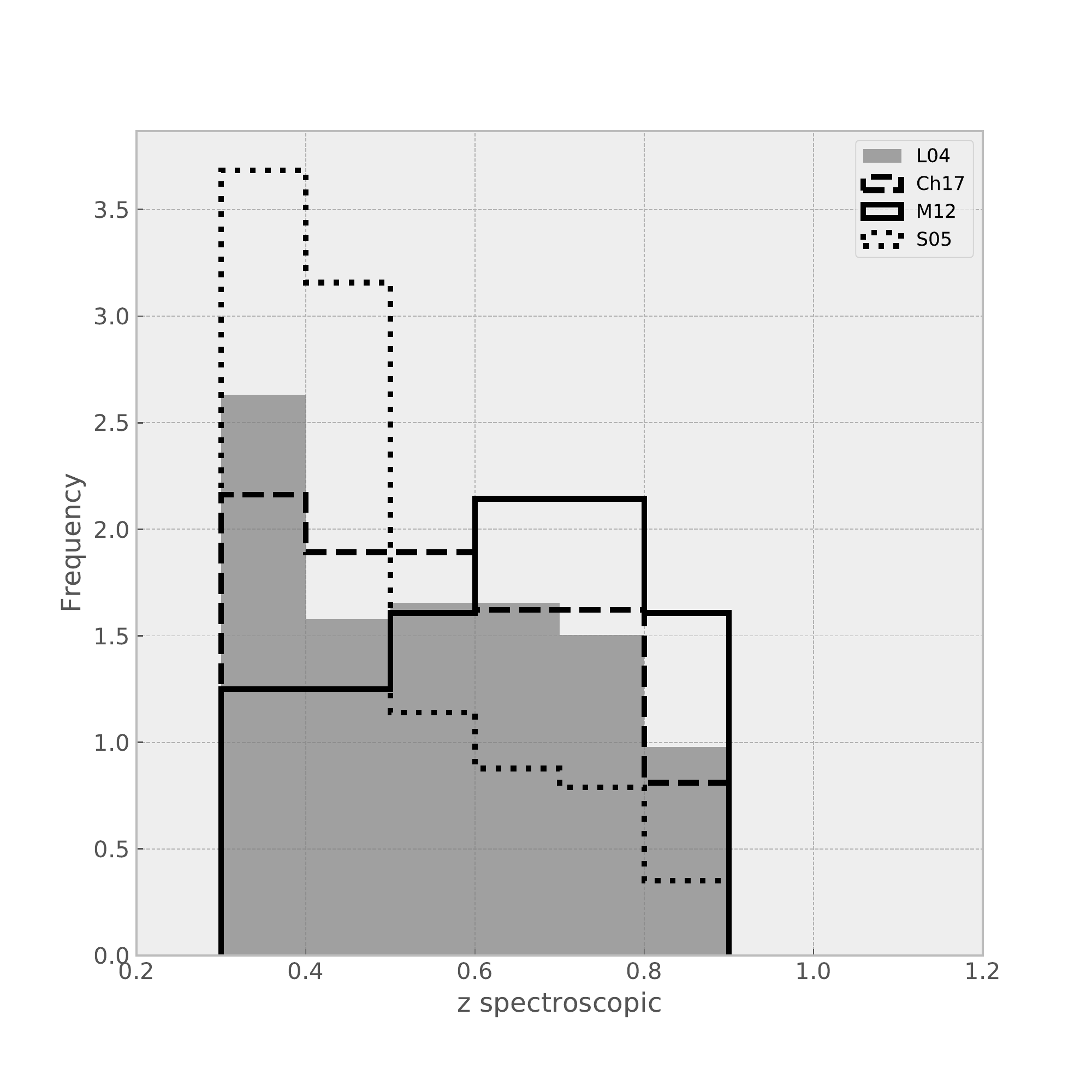}
\includegraphics[width=80mm]{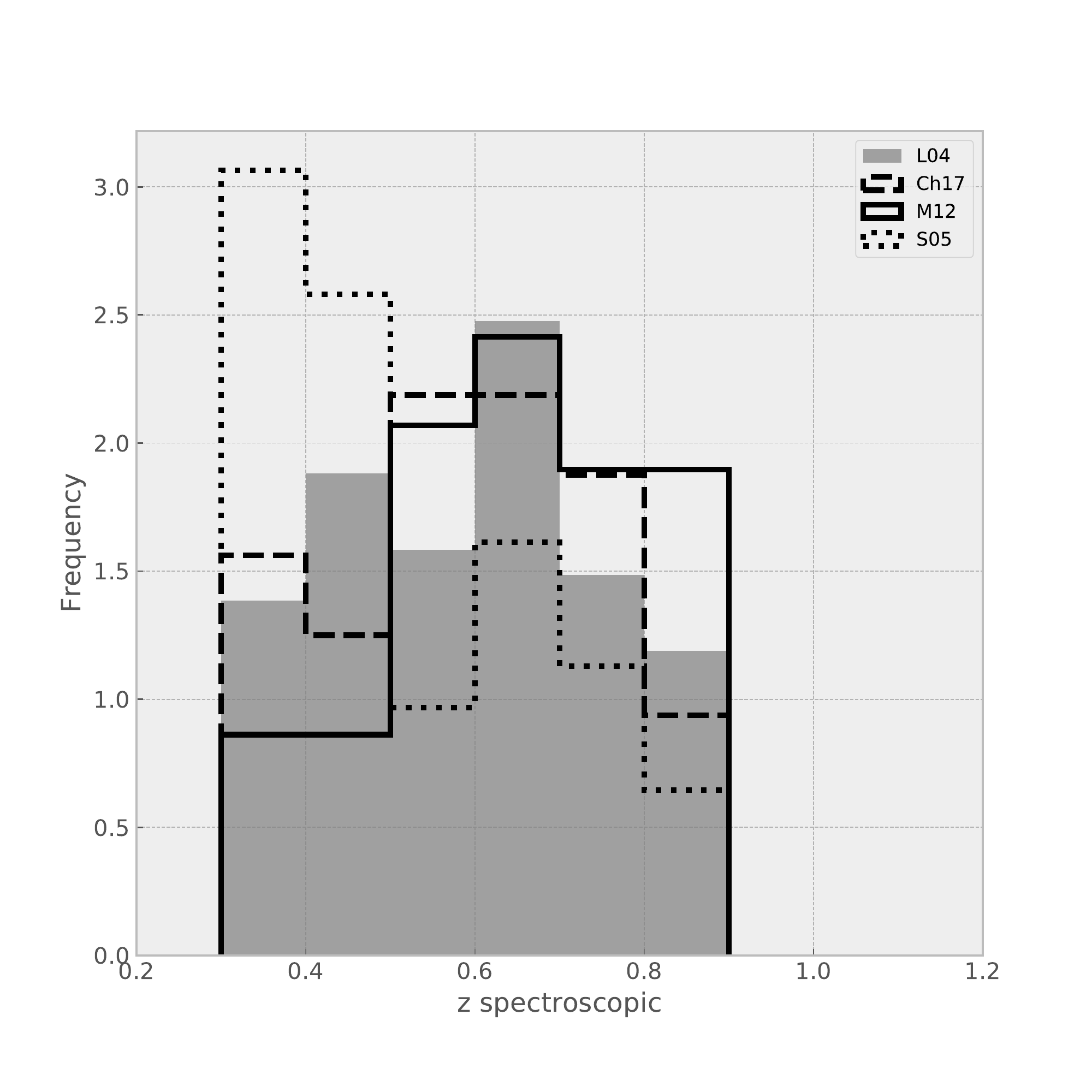}
\caption{Spectroscopic redshift distribution for AGNs selected according the different MIR and near-IR methods using the KEx (left panel) and MEx (right panel) diagrams. Point line and shaded histograms represent the distributions using the methods proposed by S05 and L04 while solid and dashed line histograms, those methods proposed by M12 and Ch17, respectively. } 
\label{zdistri}
\end{figure*}

In Figure~\ref{zdistri}, we show the spectroscopic redshift distribution for these  AGNs selected following the different MIR and near-IR methods using the KEx (left panel) and MEx (right panel) diagrams.
As it can be seen, the S05 method preferably selects AGNs at lower redshifts (z$<$0.5), according to both KEx and MEx diagrams. Up to $z=0.$9, the rest of the MIR and near-IR methods select a similar fraction of AGNs.

\section{X-ray, near-UV, optical, mid- and near-IR properties}
\label{nearuv}

\subsection{X-ray emission}

We have carried out an analysis on the X-ray emission of the pre-selected AGN sample using MIR and near-IR methods.
We cross-correlated our pre-selected AGN catalogues with the X-ray catalogue presented by \citet{civano}. The aforementioned catalogue is the COSMOS-Legacy Survey, a 4.6 Ms Chandra program on the 2.2 deg$^2$ of the COSMOS field. in Table \ref{tablex} we present the number of X-ray sources detected according to the MIR and near-IR methos of L04, Ch17, S05 and M12.
We found 14\%, 40\%, 13\% and 38\% of sources with X-ray emission according to the pre-selection methods in the MIR and near-IR of L04, Ch17, S05 and M12, respectively.

\begin{table}
\center
\caption{Number of AGN candidates with X-ray emission according to the MIR and near-IR methos of L04, Ch17, S05 and M12.}
\begin{tabular}{lcccc}
\hline
IR method & L04 &  Ch17 & S05 & M12\\
\hline
X-ray AGN candidates     & 56 &  26  & 45 & 39  \\
\hline
\label{tablex}
\end{tabular}
\end{table}

\begin{figure*}
\includegraphics[width=85mm]{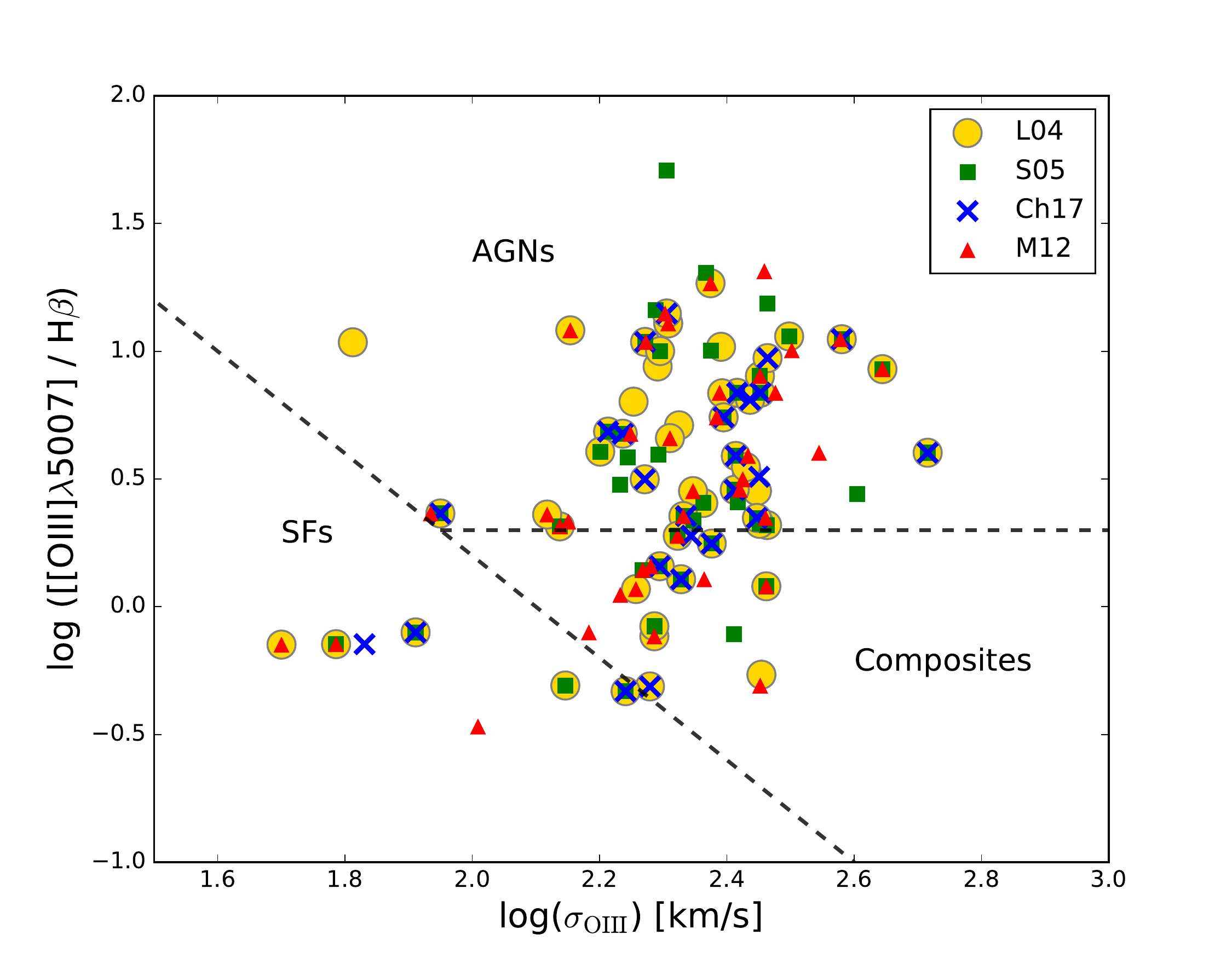}
\includegraphics[width=85mm]{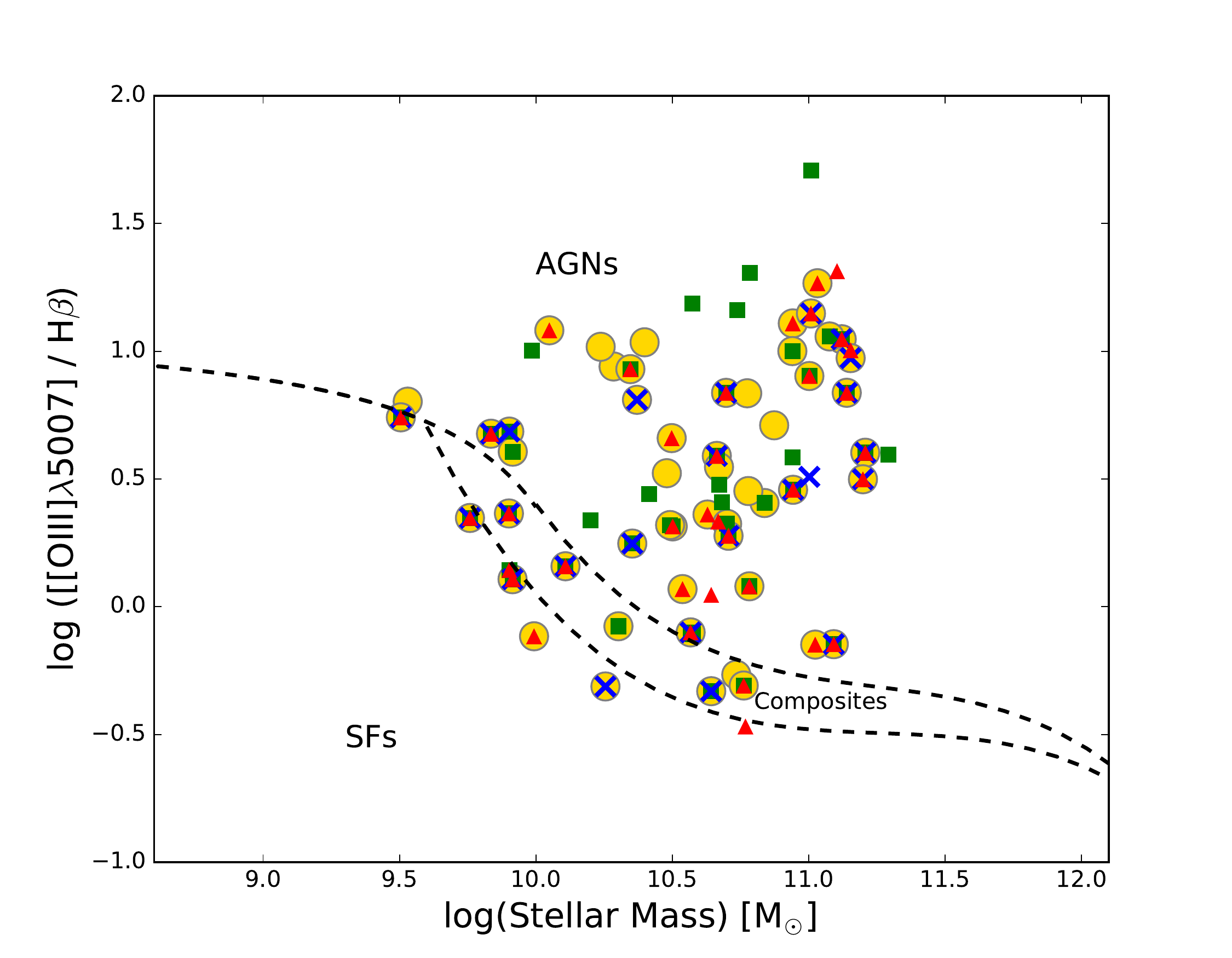}
	\caption{Sources detected with X-ray emission for AGNs pre-selected according the different MIR and near-IR methods using the KEx (left panel) and MEx (right panel) diagrams.	}
\label{xray}
\end{figure*}

In Figure \ref{xray} we plot the distribution of pre-selected AGNs with X-ray emission using the KEx (left panel) and MEx (right panel) diagrams.

It is noticeable that the success rate of the mid-IR/near-IR methods for finding objects with X-ray emission, without making any luminosity cut, is somewhat lower than the AGN percentage obtained either by the KEx and MEx methods. 
This may possibly be due to the fact that there are many obscured AGNs in our sample whose
X-ray emission is blocked by large amounts of dust (like Compton Thick AGNs).
Although it is also known that a large percentage of X-rays
 sources are located outside the selection criteria in the MIR (for example for the S05 criteria, see for instance \citet{mendez2013}.
 
The percentages of AGNs, composites and SFs with X-ray emission for each MIR and near-IR method can be found in Tables \ref{xrayK} and \ref{xrayM} using the KEx and MEx diagrams. 

\begin{figure}
    \includegraphics[width=100mm]{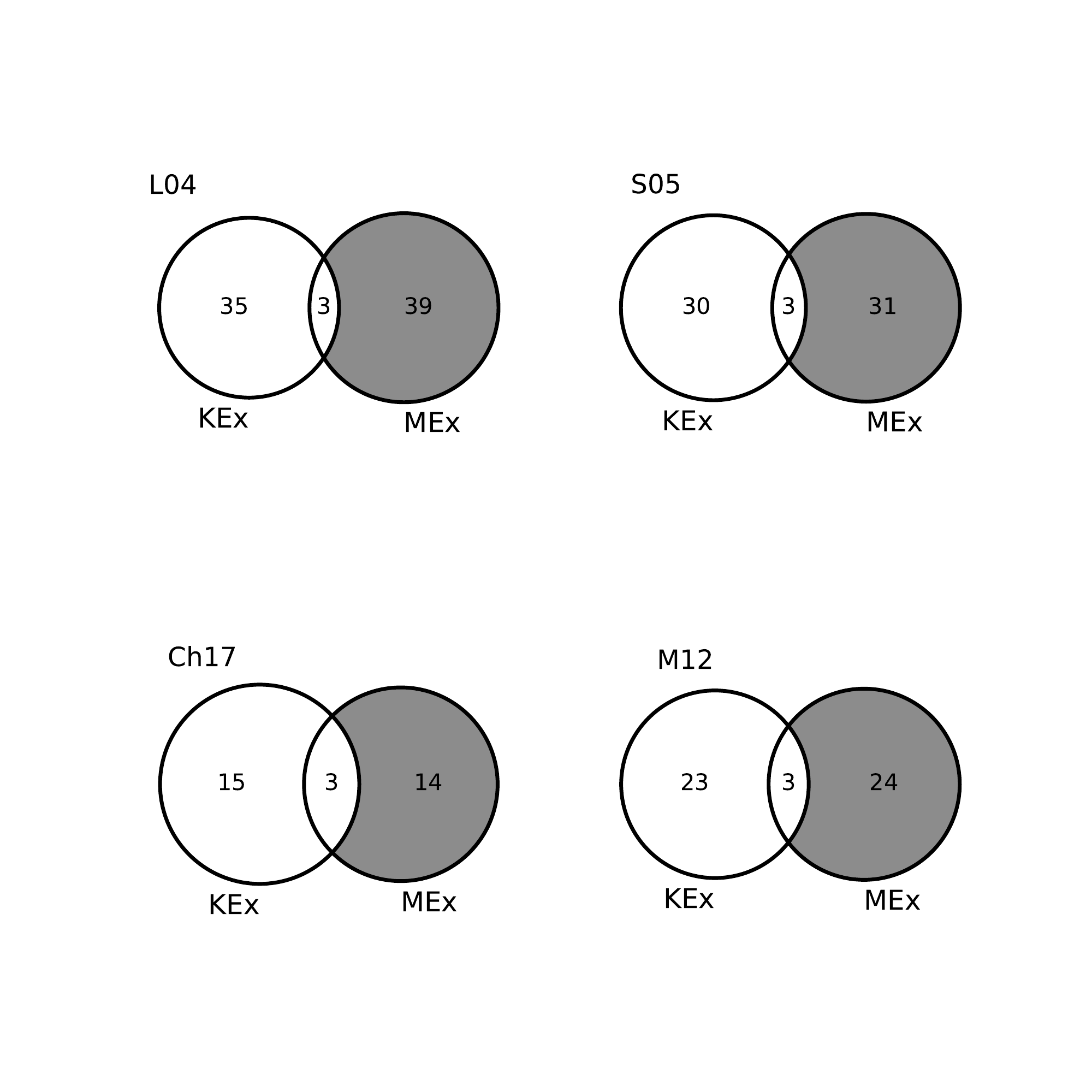}
  \includegraphics[width=100mm]{venn_xray.pdf}
	\caption{Venn diagrams showing AGNs with X-ray emission according to the L04, S05, Ch17 and M12 methods and also selected using the KEx and MEx diagrams.}
\label{vennx}
\end{figure}

In Figure \ref{vennx} we plot the Venn diagrams for the selected AGNs using the KEx and MEx diagrams with X-ray emission, according to the different selection methods in MIR and near-IR.
As can be seen, only a small percentage of sources show overlap between the different MIR methods using the KEx and MEx diagnostic diagrams.
The MIR methods with the greatest success in selecting X-ray AGNs are those of L04 and S05 ($\sim$79\%) using the MEx line diagnostic diagram.
We find that, without making any distinction between the pre-selected AGNs using MIR and near-IR methods, the MEx diagram present a higher success rate to select AGNs with X-ray emission (76\%) compared to those selected using the KEx diagram (71.7\%).

\begin{table}
\center
\caption{{Percentage of AGNs, composites and star-forming galaxies with X-ray emission selected using MIR and near-IR methods according to the KEx criteria.}}

\begin{tabular}{lcccc}
\hline
Type & L04 &  Ch17 & S05 & M12\\
\hline
AGNs       &73\% & 70\% & 73\% & 70\%  \\
Composites &18\% & 19\% & 18\% & 22\% \\
SFs         &9\% & 11\% & 9\% & 8\% \\
\hline
\label{xrayK}
\end{tabular}
\end{table}

\begin{table}
\center
\caption{{Percentage of AGNs, composites and star-forming galaxies with X-ray emission selected using MIR and near-IR methods according to the MEx criteria.}}

\begin{tabular}{lcccc}
\hline
Type & L04 &  Ch17 & S05 & M12\\
\hline
AGNs       &79\% & 71\% & 79\% & 75\%  \\
Composites &11\% & 12\% & 12\% & 8\% \\
SFs        &10\% & 17\% & 9\% & 17\% \\
\hline
\label{xrayM}
\end{tabular}
\end{table}

\subsection{Stellar mass and K$_{s}$-band absolute magnitude}

\begin{figure*}
\includegraphics[width=85mm]{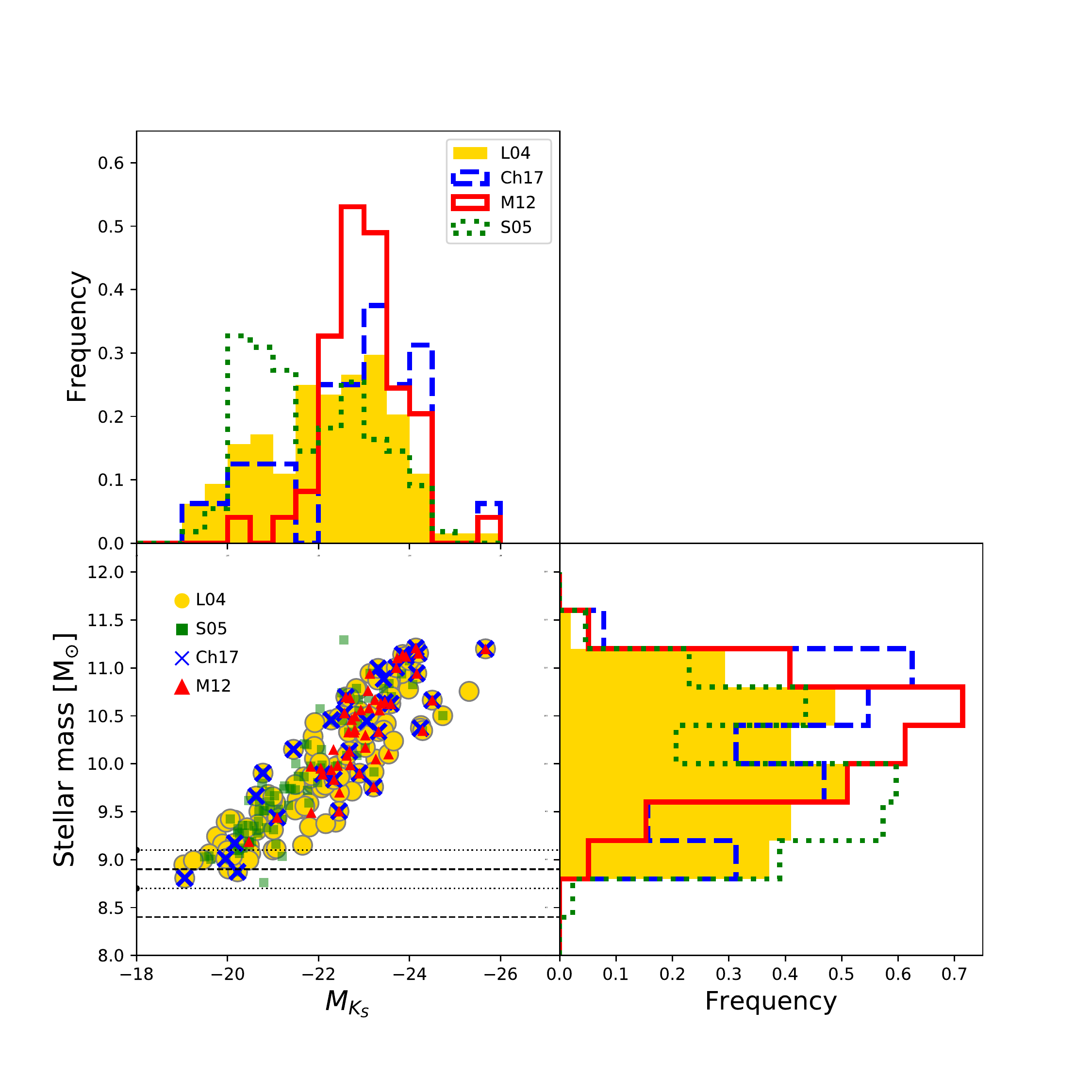}
\includegraphics[width=85mm]{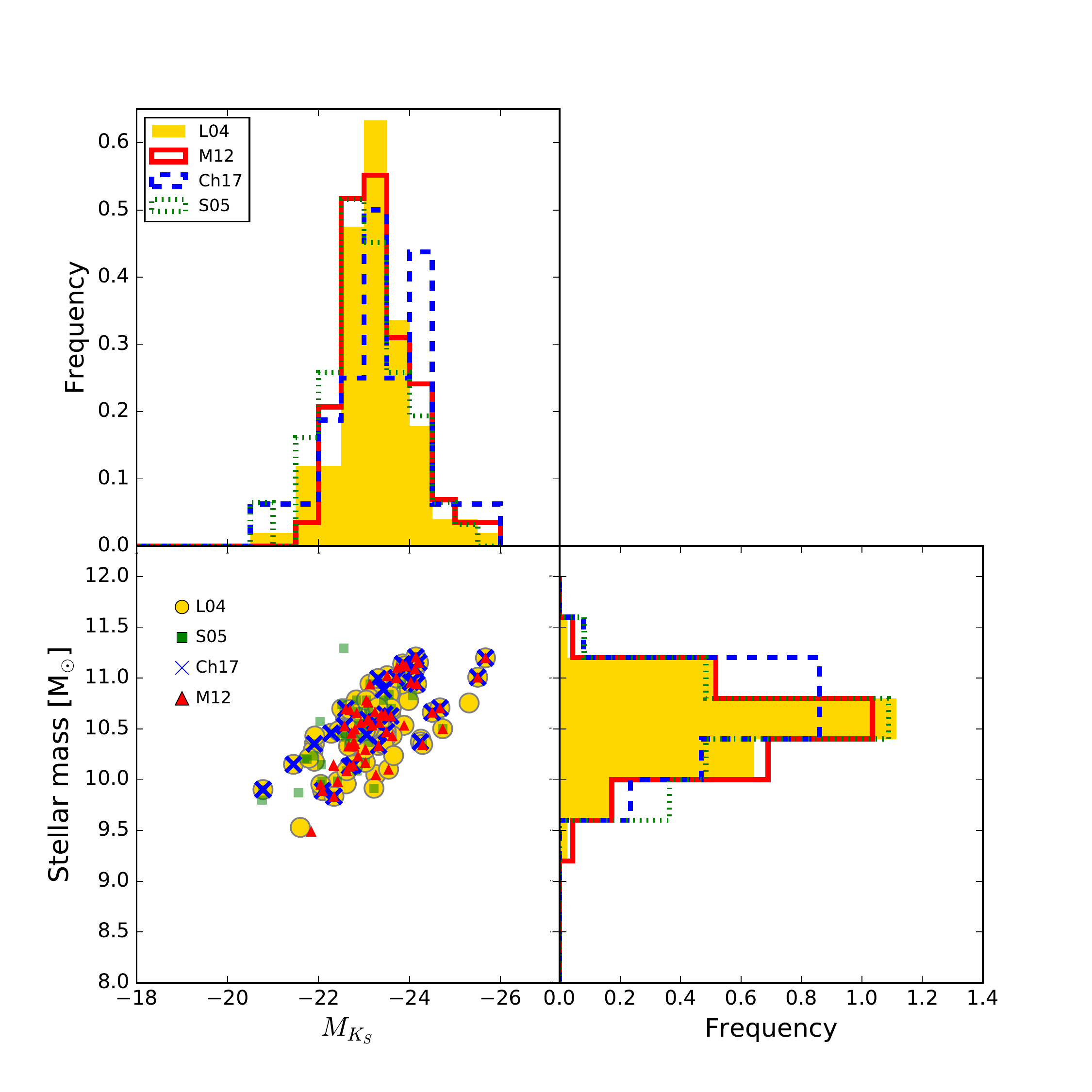}
	\caption{AGN host stellar mass vs absolute $K_s$-band magnitude using the KEx (left panel) and the MEx (right panel) line diagnostic methods. 
	The horizontal dashed line shows the stellar mass completeness obtained by \citet{laigle} for the regions named as ${\cal A}^{\rm Deep}$ (dotted lines) and ${\cal A}^{\rm UD}$ in the redshift range 0.35$<z<$0.65 and 0.65$<z<$0.95, respectively.
	The upper and right panels in each figure show the absolute $K_s$-band and the stellar mass distributions for each sample. Circles, squares, crosses and triangles represent pre-selected AGNs according to the methods of S05, L04, Ch17 and M12, respectively.}
\label{mass-lum}
\end{figure*}

In order to study the photometric properties of the MIR and near-IR selected AGNs, we analyse the stellar mass and the $K_s-$band absolute magnitude properties. We have used the MASS\_BEST estimator from the COSMOS2015 catalogue, which is estimated using the SED fitting techniques \citep{laigle}. Figure~\ref{mass-lum} shows these two parameters for AGNs selected using the KEx and MEx diagrams in the left and right panels, respectively. In each of the plots we have included AGNs pre-selected using the MIR and near-IR methods of L04, S05, Ch17 and M12. 
In Figure~\ref{mass-lum}, left panel, AGNs selected according to the S05 method present a large fraction of low-luminous and low-mass objects, compared to the other methods. The horizontal lines in this panel shows the stellar mass completeness obtained by \citet{laigle} for the regions named as ${\cal A}^{\rm Deep}$ (dotted lines) and ${\cal A}^{\rm UD}$ in the redshift range 0.35$<z<$0.65 and 0.65$<z<$0.95, respectively (see Table 6 in \citealt{laigle}). 

As can be seen at first glance, some parameters may present bimodalities in their distributions.
In order to test the existence of bimodality in the colour and/or mass distributions, we use two independent tests: a Gaussian Mixture Modeling ({\small GMM}) and the Dip tests. The {\small GMM} statistics were first implemented by \citet{muratov2010}. This code uses information from three different statistic tools: the kurtosis, the distance from the mean peaks (D), and the likelihood ratio test (LRT) in order to quantify the probability that the distributions are better described by a bimodal rather than a unimodal distribution. According to this code, the requirement for a distribution to be considered as bimodal is to obtain a negative value for the kurtosis, the separation of the peaks, D, defined as $D={\big|\mu_1-\mu_2 \big|}/\sqrt{(\sigma^2_1+\sigma^2_2)/2}$ (where $\mu_x$ 
and $\sigma_x$ are the mean and standard deviations of the two peaks of the proposed bimodal distribution), which is required to be greater than 2 and $p(\chi^2)<$0.001, a $p$-value, which gives the probability of obtaining the same $\chi^2$ from a unimodal distribution.
The Dip test was originally proposed by \citet{hartigan} and unlike the {\small GMM} test has the benefit
of being insensitive to the assumption of Gaussianity.  
The Dip test measures multimodality based on the cumulative distribution of the input sample and is defined as the maximum distance between the cumulative input distribution and
the best-fitting unimodal distribution. This
test is similar to the Kolmogorov-Smirnov test but the Dip test searches specifically for a flat step in the cumulative distribution function,
which corresponds to a “dip” in the histogram representation. 
The code was presented in \citet{muratov2010} and provides a parameter that represent the significance level at which a unimodal distribution can be rejected. \footnote{ Both {\small GMM} and Dip test codes can be download from \url{http://www-personal.umich.edu/~ognedin/gmm/}}

For the sample of pre-selected AGNs, according to the S05 method and using the KEx diagram, we find bimodality in the $M_{K_{s}}$ and stellar mass distributions, according to the following values obtained using the {\small GMM} code: kurtosis $= -1.01$, $\mu_1 = -20.71\pm0.13$, $\mu_2 = -22.70\pm0.32$, $D = 3.0\pm0.5$ and $p(\chi^2) < 0.001$ for the $M_{K_{s}}$ distribution and kurtosis $= -1.02$, $\mu_1 = 9.50\pm0.10$, $\mu_2 = 10.75\pm0.15$, $D = 3.8\pm0.4$ and $p(\chi^2) < 0.001$ for the stellar mass distribution. 
Using the Dip code we find $p=0.78$ and $p=0.83$ for the $M_{K_{s}}$ and stellar mass distributions, respectively. In both cases we have not included any sources below the stellar mass limit according to \citet{laigle}.

For the remaining objects selected from the KEx diagram and according to the methods of L04, Ch17 and M12, the two codes did not yield values corresponding to bimodal distributions for stellar mass and $M_{K_{s}}$ values.
While the stellar mass and $M_{K_{s}}$ distributions for AGNs obtained according to the MEx diagram (Figure~\ref{mass-lum}, right panel) are not bimodal for all samples of MIR and near-IR selected AGNs, according to the {\small GMM} test.

\subsection{Quiescent and star-forming galaxy samples}

\begin{figure*}
\includegraphics[width=85mm]{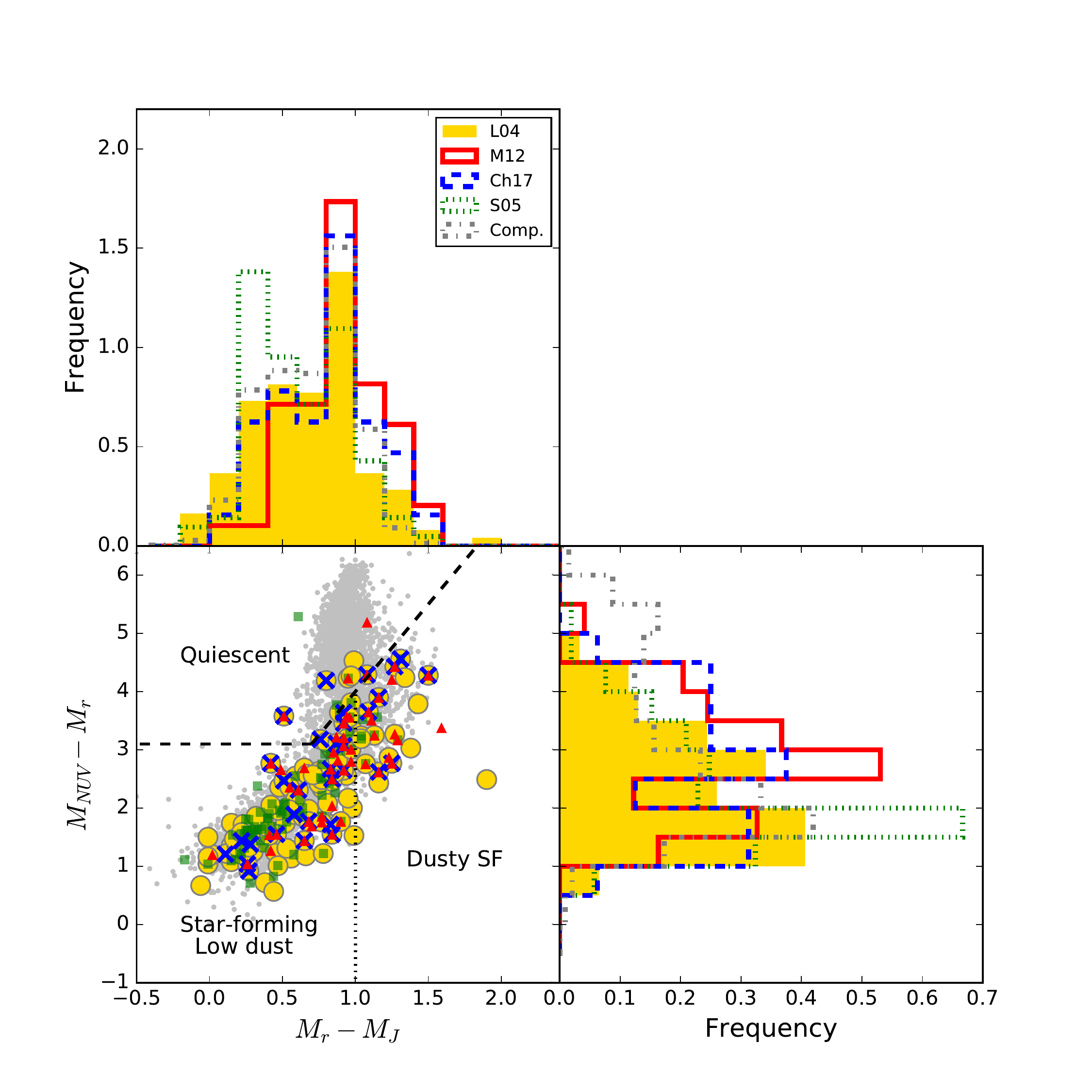}
\includegraphics[width=85mm]{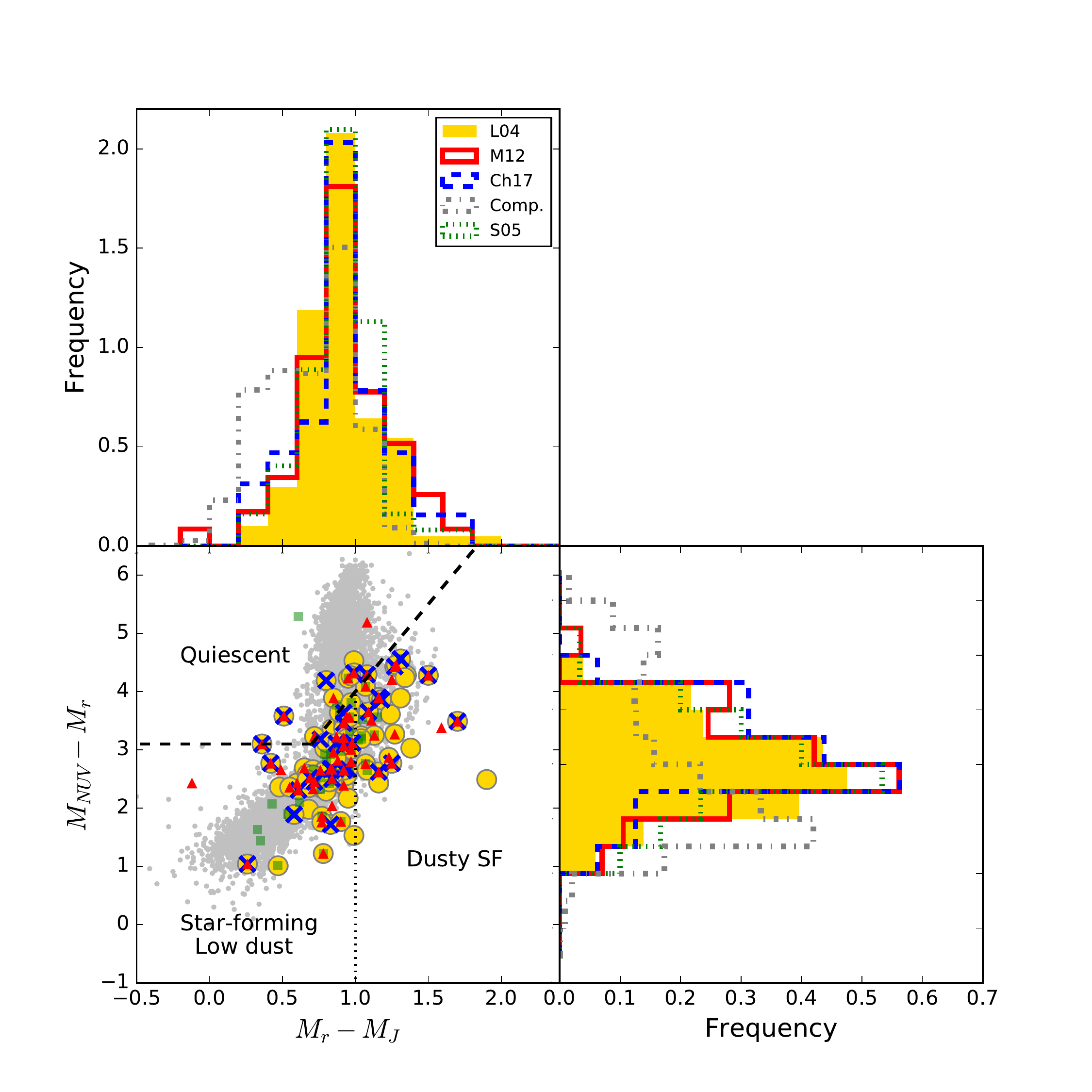}
\caption{Rest-frame ($M_{NUV}-M_r$) vs. ($M_r-M_J$) colour-colour diagram for AGN host galaxies selected according to the KEx (left panel) and the MEx (right panel) line diagnostic methods. Dashed lines mark regions which separates quiescent (upper-left corner) and star-forming galaxies. 
The symbols are the same as those used in Figure~\ref{mass-lum}.  The zCOSMOS-non-AGN sample, a comparison sample formed by galaxies with similar mass distribution at 0.3$\leq z_{\rm sp} \leq $0.9 are represented by grey points. Colour distributions are included in the upper and right panels.  The comparison sample is represented by grey dotted and circle lines. }
\label{quiescent}
\end{figure*}

\citet{williams09} employed a rest-frame colour-colour selection technique using purely photometric data to identify samples of quiescent and star-forming galaxies at redshifts $z\lesssim2$.
The quiescent samples tend to be early-type galaxies forming a different red sequence in the colour-colour observed in the $UVJ$ diagram.
Many authors have used this diagram to separate populations of predominantly blue colour star-forming galaxies from red and quiescent galaxies using a variety of rest-frame colours with similar filters \citep{patel11, arnouts13, ilbert13,straat16,fang18}.

Figure~\ref{quiescent} shows the rest-frame $M_{NUV}-M_r$ vs $M_r-M_J$ colour-colour diagrams for AGNs pre-selected using the L04, S05, Ch12 and M12 methods for AGNs obtained from KEx and MEx diagrams (left and right panels, respectively). The dashed lines represent the boundaries that separate regions occupied by quiescent and star-forming galaxies taken from \citet{ilbert13}. In each figure, we have also included the corresponding colour distributions for each MIR and near-IR methods (upper and right panels). We have also included a sample of galaxies taken from zCOSMOS-$good$ with 0.3$ \leq z_{\rm sp} \leq $0.9 and without any pre-selected AGNs according to L04, S05, Ch17 and M12 methods (grey dots and dotted histograms). Hereafter we will call this sample as zCOSMOS-non-{\footnotesize AGN}.

\citet{spitler14} studied the high redshift massive galaxy population in the ZFOURGE survey. They have used an $U-V$ vs. $V-J$ colour-colour diagram in order to analysed the quiescent and star-forming galaxy populations. They have split the star-forming sample into two groups: one with large dust content ($V-J > 1.2$) and other with low dust content ($V-J < 1.2$).
We have used this separation criterion transforming the colours $V-J$ into $R-J$ through a linear relationship found between both colours using our zCOSMOS-$good$ sample:
$$(M_r-M_J)=(0.82\pm0.01)\times(M_V-M_J)-(0.080\pm0.002)$$.

\begin{figure*}
  \includegraphics[width=200mm]{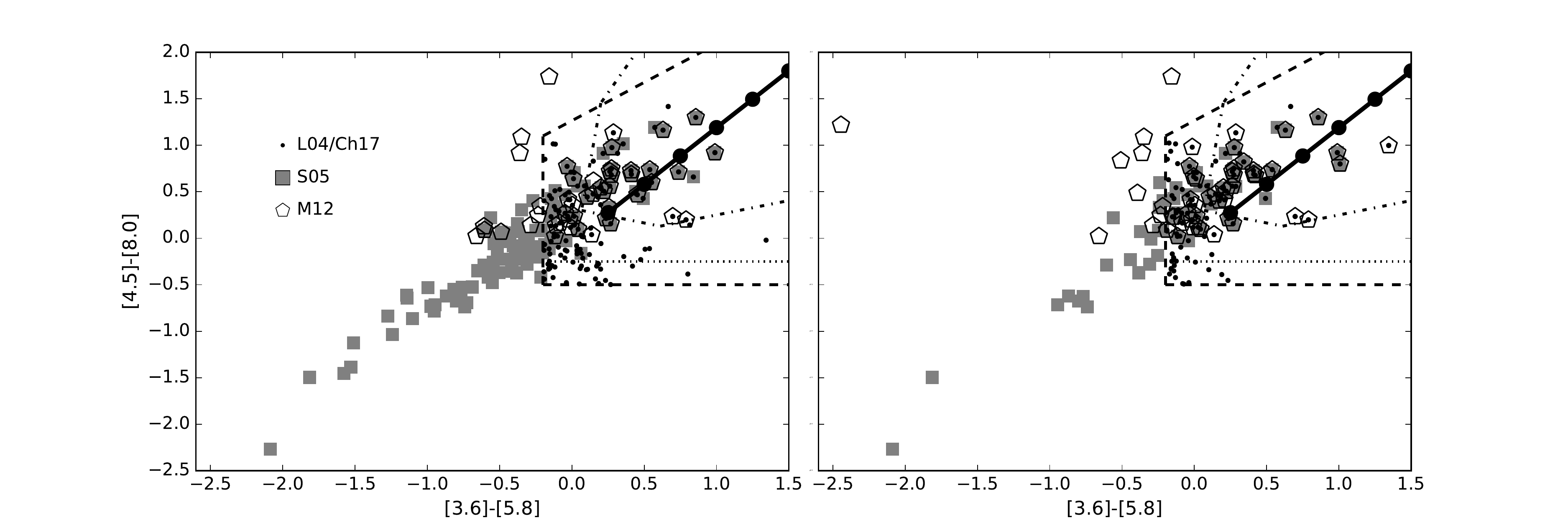}
\caption{Mid-IR ($[4.5]-[8.0]$ vs $[3.6]-[5.8]$) colour-colour diagrams for AGNs selecting according the methods present by L04 and Ch17 (small dots). We have included also the AGN colours using the methods present by S05 (squares) and M12 (pentagons). The region marked with dashed and dot-dashed lines show the selection criteria of L04 and Ch17, respectively. Horizontal dot lines shows a correction of $+0.25$ magnitudes proposed by \citep{L07}. 
The line with filled circles in black indicates the locus of sources whose spectrum can be described as a power-law with $\alpha$ = $-$0.5 (lower left) to $\alpha$ = $-$3.0 (upper right).}
\label{agnK}
\end{figure*}

The majority of the AGNs selected by the IR methods are located within the locus where the star-forming galaxies reside. 
The S05 sample presents a considerable fraction of blue objects located in the region populated by SF galaxies with low dust content in both $M_{NUV}-M_r$ and $M_r-M_J$ colours, for the sample of AGNs selected according to the KEx diagram. On the other hand, AGNs selected according to the MEx diagram do not present a locus of objects in the area occupied by low dust content star-forming galaxies as observed in the KEx diagram.
This excess of sources with low dust content observed in the methods of L04 and S05 according to the KEx diagram is evidenced by the large number of objects that do not present overlap as we have shown in Figure \ref{venn}.

Without making any distinction between AGNs pre-selected using MIR and near-IR methods, we find that only 5.4\% and 11.5\% of AGNs are located in the quiescent region, according to the KEx and MEx diagrams, respectively. 

Within the observed bimodality in galaxy colours, some authors affirm that there would be a transition population between the red and the blue populations, the so-called the Green Valley galaxies \citep{wyder07,salim07, salim14}. This region is thought to represent a transition in the life of galaxies ranging from blue galaxies with high star formation rates to passive galaxies with predominantly red colours \citep{salim07, salim14}. 
According to some authors the limit values that define this region are between 3$<M_{NUV}-M_r<$5 \citep{salim07}, 3.5$<M_{NUV}-M_r<$4.5 \citep{salim2009}, or
3.2$<M_{NUV}-M_r<$4.1 \citep{mendez2011} for galaxies at 0.4$<z<$1.2, which coincides for a sample of galaxies at low redshifts studied by \citet{coenda2018} with similar host stellar masses. 
In the left panel of Figure~\ref{quiescent}, we find that the 
colours of the AGN samples selected using the KEx diagram are located mostly in the blue branch of the colour-colour diagram, while the sample of AGNs selected using the MEx diagram are located near the green valley although with little bluer colours than the average position of green valley galaxies.

The colour distribution for the zCOSMOS-non-{\footnotesize AGN} sample shows a clear bimodality. According to the {\small GMM} code we have found the following values: 
kurtosis $= -0.888$, $\mu_1 = 1.9\pm0.03$, $\mu_2 = 4.60\pm0.06$, $D = 3.43\pm0.11$ and $p(\chi^2) < 0.001$, while using the Dip test, we find $p=1.0$. 
For the sample of AGNs selected using the KEx diagram, we find that the $M_r-M_J$ colour distributions according to the Ch17 and M12 are not bimodal, while the corresponding colour distribution of S05 is bimodal according to the {\small GMM and Dip codes}. We have obtained the following values using the {\small GMM} code: kurtosis $= -0.68$, $\mu_1 = 0.35\pm0.05$, $\mu_2 = 0.92\pm0.10$, $D = 3.0\pm0.6$, $p(\chi^2) < 0.001$ and $p=0.9$. The sample of L04 present a moderate bimodality according to the $p$-value $p(\chi^2)=0.22$ and $p=0.15$, using the {\small GMM} and Dip codes, respectively. 
For the $M_{NUV}-M_r$ colour distribution the results obtained with the {\small GMM} code shows that the S05 method is not bimodal, while the M12 and Ch17 methods present a moderately bimodality ($p(\chi^2)=0.117$ and $0.09$, and $p=0.31 $ and $p=0.1$, using the {\small GMM} and Dip tests respectively) and the L04 method presents bimodal colour distribution with the following values: kurtosis $= -0.68$, $\mu_1 = 1.43\pm0.10$, $\mu_2 = 2.9\pm0.15$, $D = 2.3\pm0.3$ and $p(\chi^2) < 0.001$ using the {\small GMM} code. Using the Dip test, we have obtained $p = 0.76$.
For the case of the AGNs selected using the MEx diagram (right panel of the Figure~\ref{quiescent}) all the colour distributions are well represented by unimodal distributions.
For the $M_r-M_J$ colours the distributions are well characterised with $\mu = 0.91\pm0.01$ and $\sigma$ = 0.28$\pm$0.04 while for the $M_{NUV}-M_r$ colour distribution, we find $\mu = 3.0\pm0.1$ and $\sigma$ = 0.83$\pm$0.03.

\begin{figure*}
\includegraphics[width=85mm]{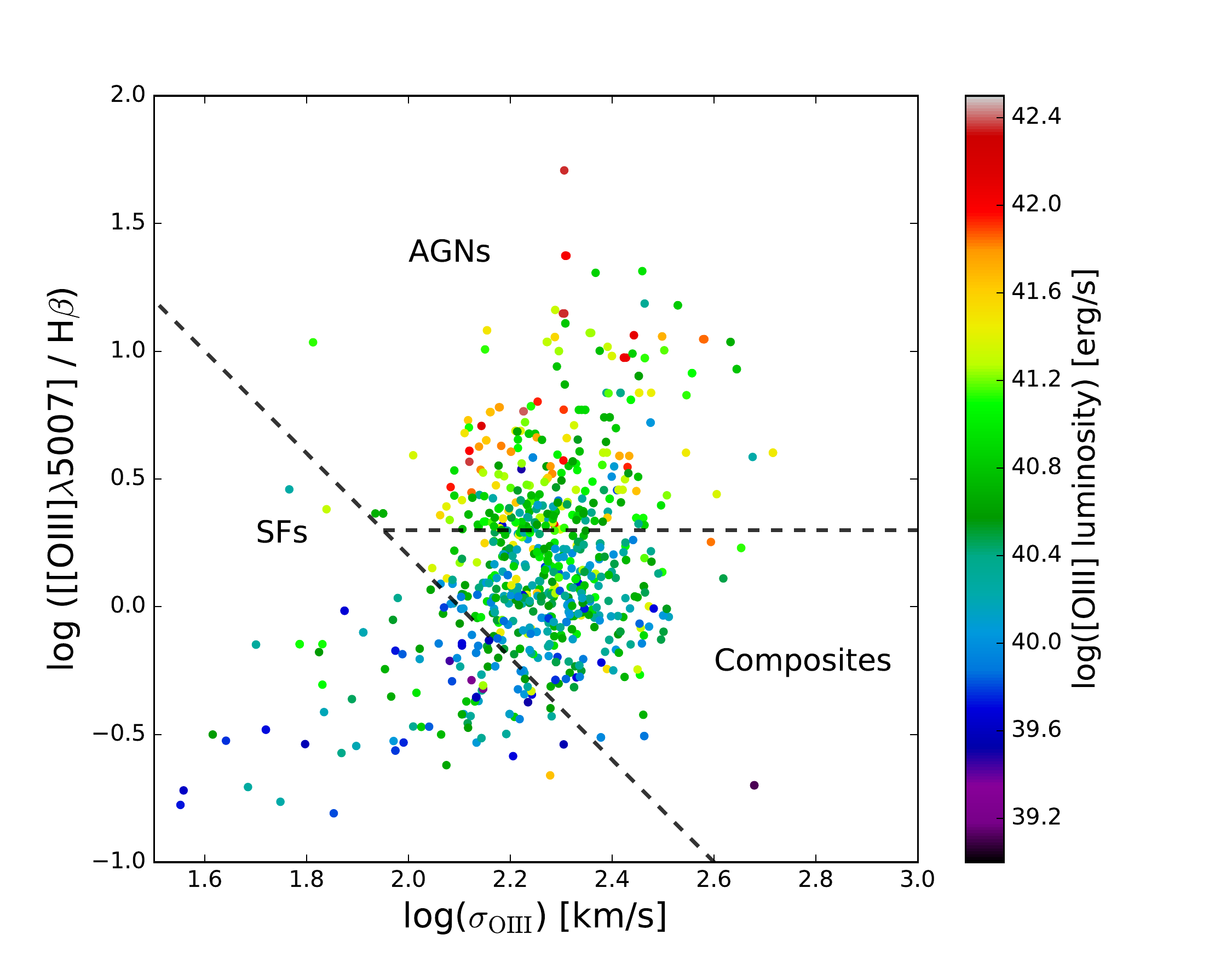}
\includegraphics[width=85mm]{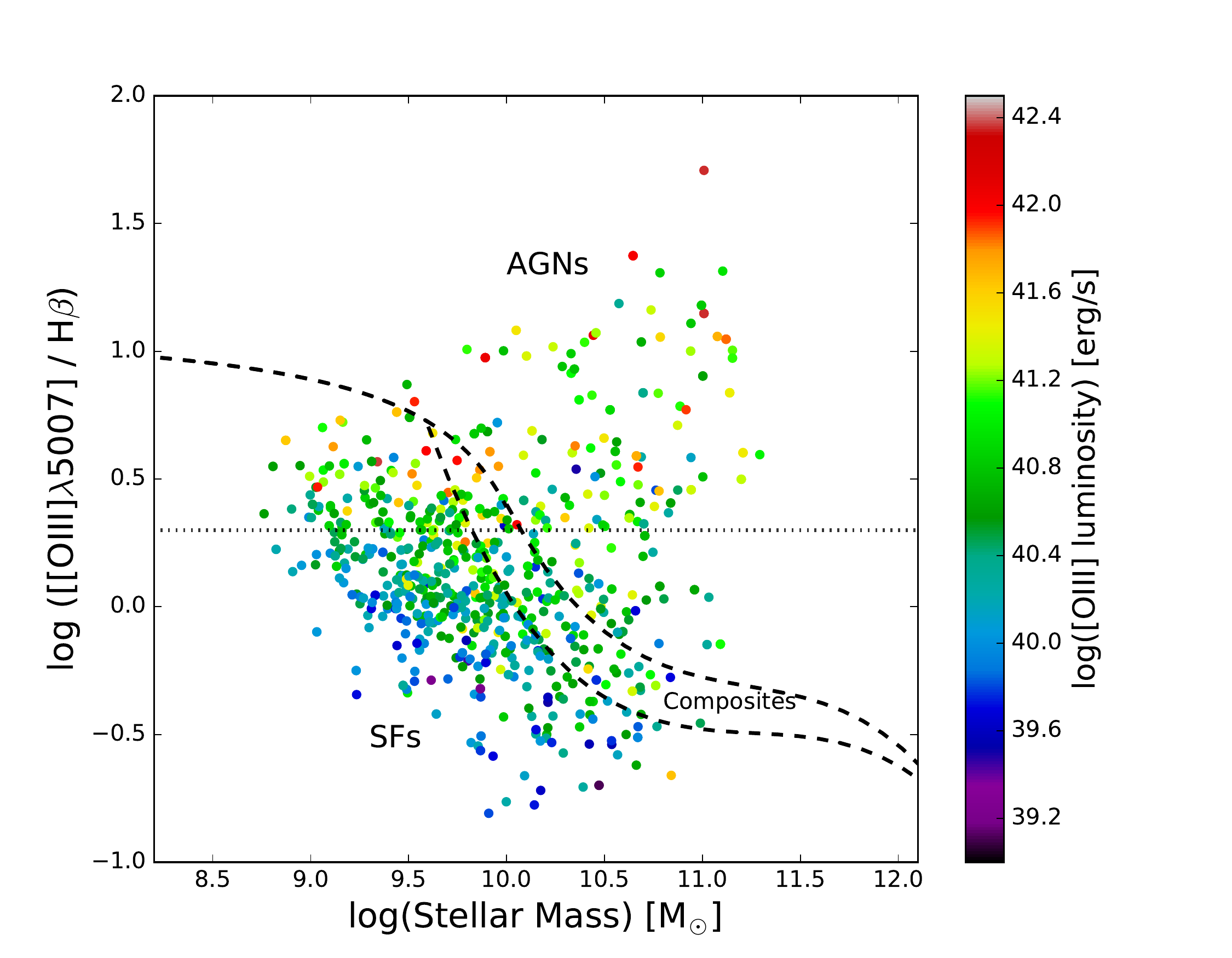}
\caption{The kinematics (left panel) and mass-excitation (right panel) diagrams for sources selected according to MIR and near-IR methods proposed by S05, L04, Ch17 and M12. The vertical bar represent the [O III] $\lambda$5007 luminosity of each source. The horizontal dotted line in the right panel shows the limit at y=0.3 that separate AGNs in the KEx diagram.}
\label{OIII}
\end{figure*}

\begin{figure*}
\includegraphics[width=85mm]{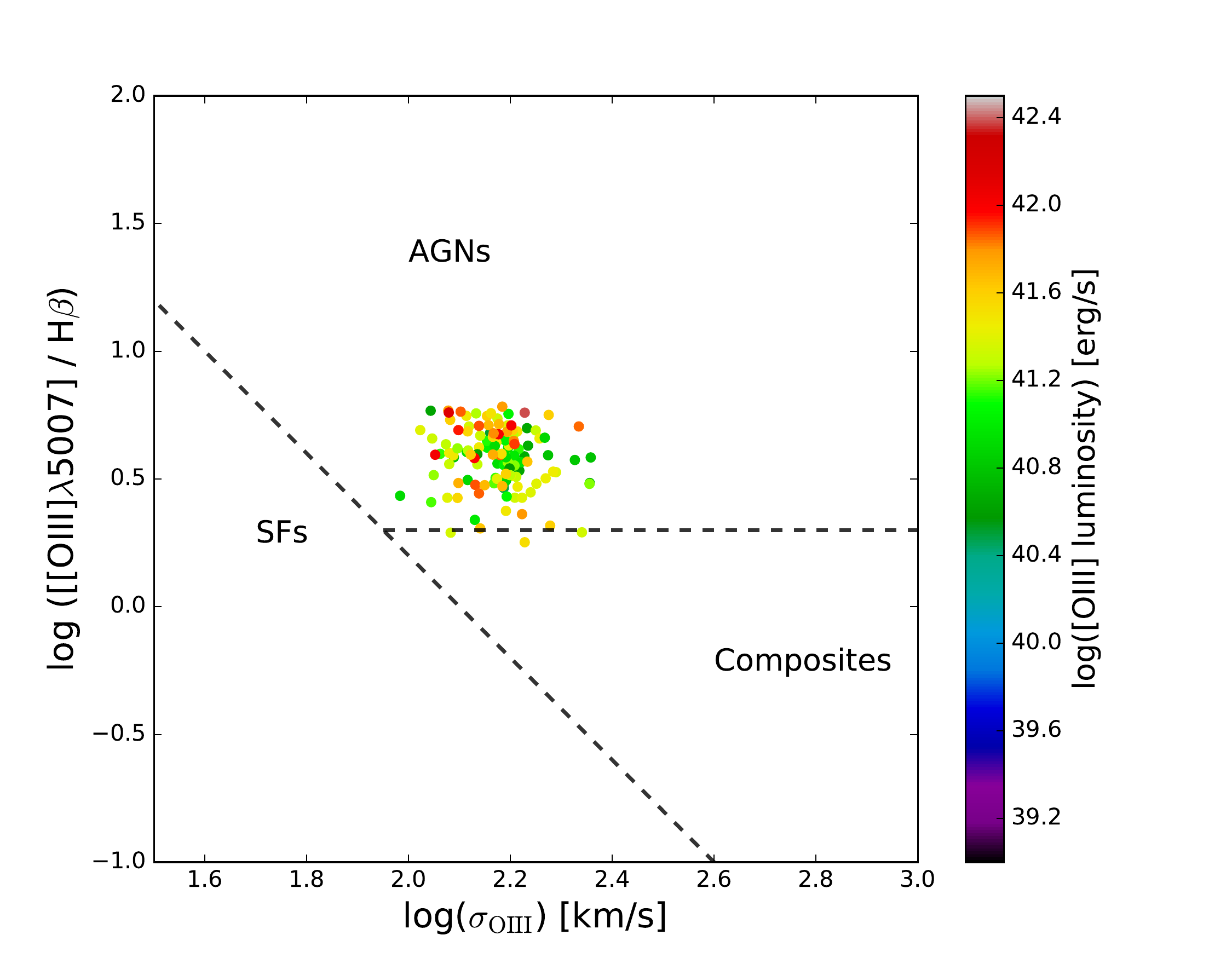}
\includegraphics[width=85mm]{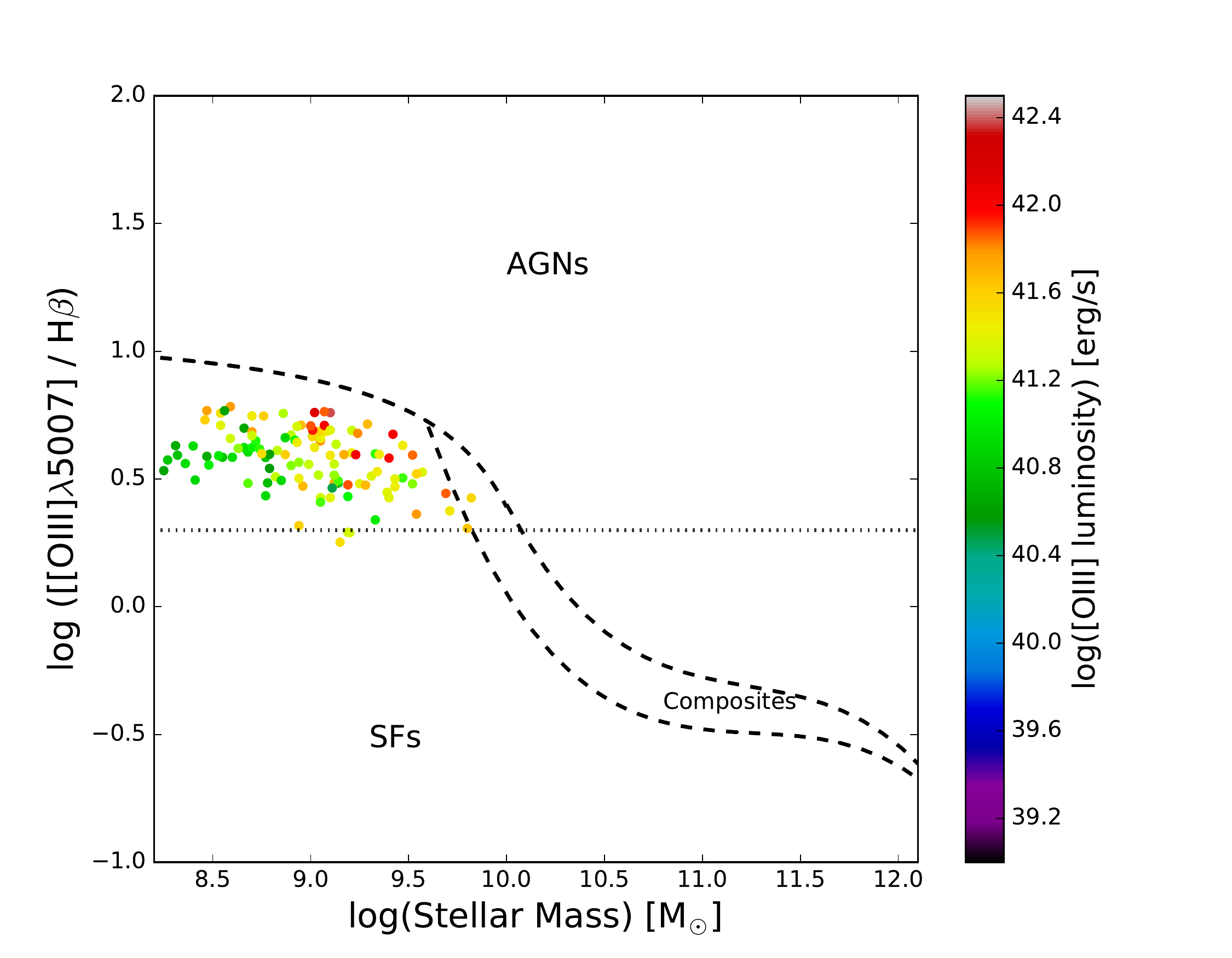}
\caption{The kinematics (left panel) and mass-excitation (right panel) diagrams for extreme emission-line star-forming galaxies taken from \citet{amorin15}. The vertical bar represents the [O III] $\lambda$5007 luminosity of the sources. The horizontal dotted line in the right panel shows the limit at y=0.3 that separates AGNs in the kinematics-excitation diagram.}
\label{amorin}
\end{figure*}

\subsection{AGN MIR colour-colour diagram}

In this section we study the MIR IRAC four colour position of each AGN according to the MIR and near-IR methods.  
In Figure~\ref{agnK}, we plot the MIR colour-colour diagrams for pre-selected AGNs according to the methods present by L04 and Ch17 (small dots). We have included also in the same plot, the corresponding $[4.5]-[8.0]$ and $[3.6]-[5.8]$ AGN candidate colours using the methods present by S05 (squares) and M12 (pentagons). The region marked with dashed and dot-dashed lines show the selection criteria of L04 and Ch17, respectively.  
Horizontal dot lines show a correction of $+0.25$ magnitudes proposed by \citep{L07}. These authors presented a small modification to the AGN selection box, moving the log(S$_{5.8}$/S$_{3.6}$) cut by $+0.1$ compared to \citet{L04} in order to remove possible non-AGN contaminants. As these authors claim the exact position of this cut is not critical and the line diagnostics made through KEx and MEx diagrams eliminates other possible contaminants.  
We have calculated the percentages of AGNs, Composites and SF galaxies of the objects within the rectangle defined by the difference between the selection boxes of L04 and L07 (see Figure \ref{agnK}). According to the KEx diagram we find (AGNs, Composites, SF) = (30,48,22)\% and using the MEx diagram (AGNs, Composites, SFs) = (19,16,65)\%. 
We find that the percentages of the different galaxy types are similar to those found by L04 (see Tables 1 and 2).
Due to this, in this work we use the criterion of L04 instead of the one presented in \citet{L07}.
We have also plot a solid line that represents the power-law locus, i.e., the line on which a source with a perfect power-law SED would fall. Filled circles along this line denote power-law slopes from $\alpha = -$0.5 (lower left) to $\alpha = -$3.0 (upper right).
As can be appreciated some dots corresponding to the S05 criterion fall outside the L04 and Ch17 boxes. These have bluer colours in both $[3.6]-[5.8]$ $[4.5]-[8.0]$ and are on average faint in the $K_s-$band compared to the rest of the other MIR and near-IR methods.
These sources seem to continue the power-law towards positive values of $\alpha$. As stated in \citet{barmby06}, AGN-dominated objects would have red
power laws with $\alpha < 0$ (preferentially with $\alpha < -0.5$, \citealt{donley07, donley08}), while stellar-dominated galaxies would have blue SEDs with $\alpha \sim +2$. 
This evidence together with the results found from Figures~\ref{mass-lum} and~\ref{quiescent} indicates that some of the objects selected according to the method proposed by S05 represent low mass faint star-forming (with a possible weak AGN component) galaxies at lower redshifts.


\section{Properties of AGNs selected by KEx and MEx diagnostic diagrams}
\label{6}

\subsection{[OIII] line properties} 

In this section we will study the AGN properties according to the KEx and MEx diagrams without making a distinction between the pre-selected AGNs using MIR and near-IR methods.
In Figure~\ref{OIII} we show the KEx and MEx diagrams for the sample of MIR and near-IR pre-selected AGNs using the methods of L04, S05, Ch17 and M12. 
The vertical bar indicates the \oiii\ luminosity values for each AGN. 
We calculated the \oiii\ luminosity using the standard formula:

\begin{eqnarray}
L_{\rm \oiii}=\frac{4\pi d^2_{L}}{(1+z)}f_{\rm \oiii} 
\end{eqnarray}

where $d_{L}$ is the luminosity distance and $f_{\rm \oiii}$ is the \oiiil\ line flux.

Due to the redshift range of the AGN sample and the spectral coverage 5550-9450 \AA\ of the VIMOS data that we use in this work, we have not corrected the \oiiil\ flux to account for the absorption due the narrow-line region itself (see for instance \citealt{maiolino98, bassani98, vignali06, panessa06, lamastra09, vignali10}). This reddening correction uses the H${\alpha}$/H${\beta}$ Balmer decrement and we have only H$\alpha$ emission up to z$\sim$0.4.

It can be seen that objects in the KEx diagram (left panel of Figure~\ref{OIII}) located in the area where AGNs reside have higher \oiii\ luminosities compared to objects located in areas populated by star-forming and composite galaxies.
In the right panel of Figure~\ref{OIII}, we plot the MEx diagram for AGNs pre-selected in the MIR and near-IR wavelengths. We have also included an horizontal line at log(\oiii/\hb)=0.3 which represents the 
limit for the vast majority of AGNs selected using the KEx diagram.
According to this diagram, we find that some objects located in the region populated by star-forming galaxies, those around the coordinate (log({Stellar~mass}),log(\oiii/\hb))=$(9.4,0.6)$ have also high \oiii\ luminosities compared with those found in the AGN sample. It is known that the \oiii\ emission in AGNs might come from the narrow-line region, but also from the HII regions of SF galaxies.
As postulated by \citet{Zh18}, the \oiii\ emission from AGNs traces the bulge kinematics, while in star-forming galaxies comes from HII regions located along the disk.

In order to establish whether these objects are AGNs as determined by the KEx diagram or star-forming galaxies with high \oiii\ luminosities according to the MEx diagram, we use a sample of star-forming galaxies with extreme emission lines taken from \citet{amorin15}.  These authors studied a sample of 183 extreme emission-line galaxies (EELGs) at redshift 0.11 $\lesssim$ z $\lesssim$ 0.93 selected from the 20k zCOSMOS survey with unusually large emission line equivalent widths and high specific star formation rates. These objects were identified with compact, low-mass, low-metallicity, vigorously star-forming systems associated with luminous and higher-z counterparts of nearby HII galaxies and blue compact dwarfs. In order to elucidate this difference, in Figure~\ref{amorin}, we plot the KEx (left panel) and MEx (right panel) diagrams for a sample of star-forming galaxies with extreme emission lines with $0.3 \leq z_{\rm sp} \leq 0.9$ taken from \citet{amorin15}. 
For the case of the KEx diagram we have measured the width of the \oiiil\ emission line ($\sigma_{\rm \oiii}$) as described in Section 3.
The vertical bar represents the \oiii\ luminosities with the same scale as in Figure~\ref{OIII}.
It can be seen in the right panel of Figure~\ref{amorin},  that all these SF galaxies have log(\oiii/\hb)$ > $0.3 and high \oiii\ luminosity values, some even larger than that observed for AGNs in Figure~\ref{OIII}.
We can see that all SF forming galaxies selected in the \citet{amorin15} catalogue are located within the boundaries of the SF galaxies according to the MEx diagram. However, it can be seen that the same sample of SF galaxies are found within the region populated by AGNs according to the KEx method. 
This shows that the width of OIII used in the KEx diagram does not allow a clear separation between the emission of the \oiiil\ lines coming from the AGN from that originating in the HII regions of galaxies with high star formation rates.

\section{X-ray emission, black hole mass and morphology of the AGNs selected through the MEx diagram}

We have shown that the KEx diagram presents great contamination by SF
galaxies within the limits that demarcate the
selection of AGNs.  In this Section,  we analyse the AGN properties selected
only according to the MEx diagram.

\subsection{AGN X-ray emission properties}

 We investigate here, the X-ray properties of AGNs.
The hardness ratio (HR) is an indication of the spectral shape and can be used to separate obscured and unobscured AGNs in the X-ray wavelengths.
For this reason we use the HR measurements from the catalogue of \citet{civano}, which is defined as,

\begin{eqnarray}
HR=\frac{H-R}{H+R},
\end{eqnarray}

where $H$ and $S$ are the count rates in the hard and soft bands, respectively.

\begin{figure}
  \includegraphics[width=85mm]{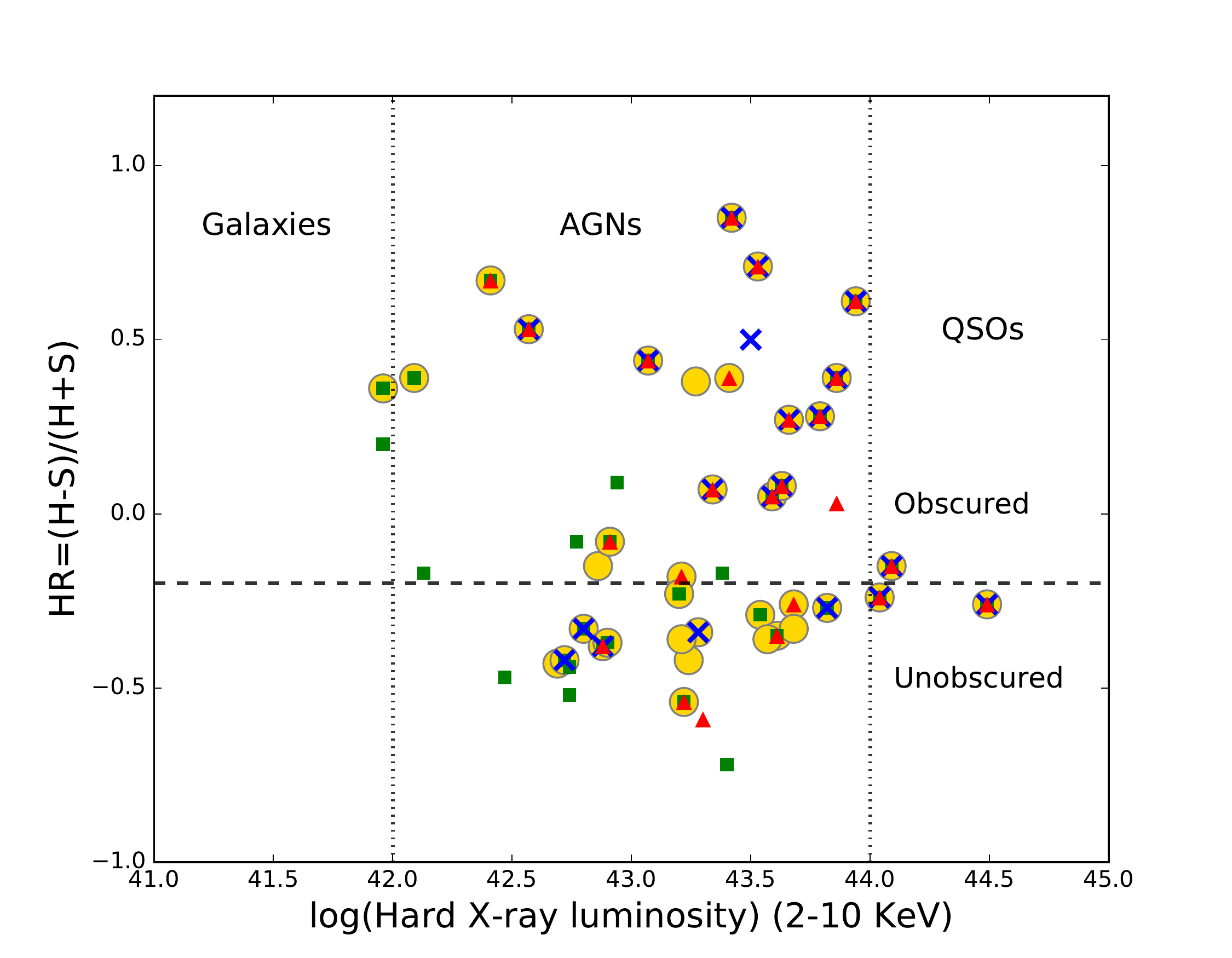}
\caption{Hardness ratio as a function of hard (2-10 keV)
X-ray luminosity. Vertical dashed lines show the typical separation for normal galaxies, AGNs and quasars used in the X-rays. The symbols are the same as those used in Figure 4.
}
\label{hr}
\end{figure}

In Figure \ref{hr}, we plot the HR as a function of rest-frame hard X-ray luminosity for  each  MIR  and  near-IR method using the MEx diagram which is calculated as,

\begin{eqnarray}
 L_{X} = 4 \pi\,d^{2}_{L}\,f_{X}\,(1+z)^{\Gamma-2}\,\rm{erg}\,\rm{s}^{-1}, 
\end{eqnarray}

where $d_{L}$ is the luminosity distance (cm), \emph{f}$_{X}$ is the X-ray flux (erg s$^{-1}$ cm$^{-2}$) in the hard-band and the photon index was assumed to be $\Gamma=1.8$ \citep{Tozzi06}. 

The dashed horizontal line shows the HR value (HR$=-$0.2) for a source with a neutral hydrogen column density, N$_H$ > 10$^{22}$ cm$^2$ \citep{civano2012}, which is used by several authors \citep{gilli09, treister09, marchesi} to separate obscured and unobscured sources in the X-rays at all redshifts.

Vertical dashed lines show the typical separations used in the X-rays following \citet{treister09} for normal galaxies (L$_X <$ 10$^{42}$ \ergs), AGNs (10$^{42} <$ L$_X <$ 10$^{44}$ \ergs) and quasars (L$_X >$ 10$^{44}$ \ergs).

For each MIR and near-IR method, we find the percentage of obscured (HR $> -$0.2) sources of 52, 54, 65 and 71\% according to the methods proposed by L04, Ch17, S05 and M12, respectively.
Without making distinctions between AGNs pre-selected using MIR and near-IR methods, we find 59.13\% and 40.87\% of AGNs obscured and unobscured, respectively.
We have also performed the same calculations for the sample of X-ray sources taken from \citet{civano} with $0.3\leq z_{\rm sp}\leq 0.9$. In this sample, we find very similar results of obscured (50.6\%) and unobscured (49.4\%) sources. 
By comparison, we have found that our AGN samples selected according the S05 and M12 methods, also identified in the MEx diagram, have a 15 to 21\% excess of obscured objects (HR $>-$0.2).

\subsection{Black Hole and AGN accreting properties}
\label{bh}

\begin{figure}
  \includegraphics[width=85mm]{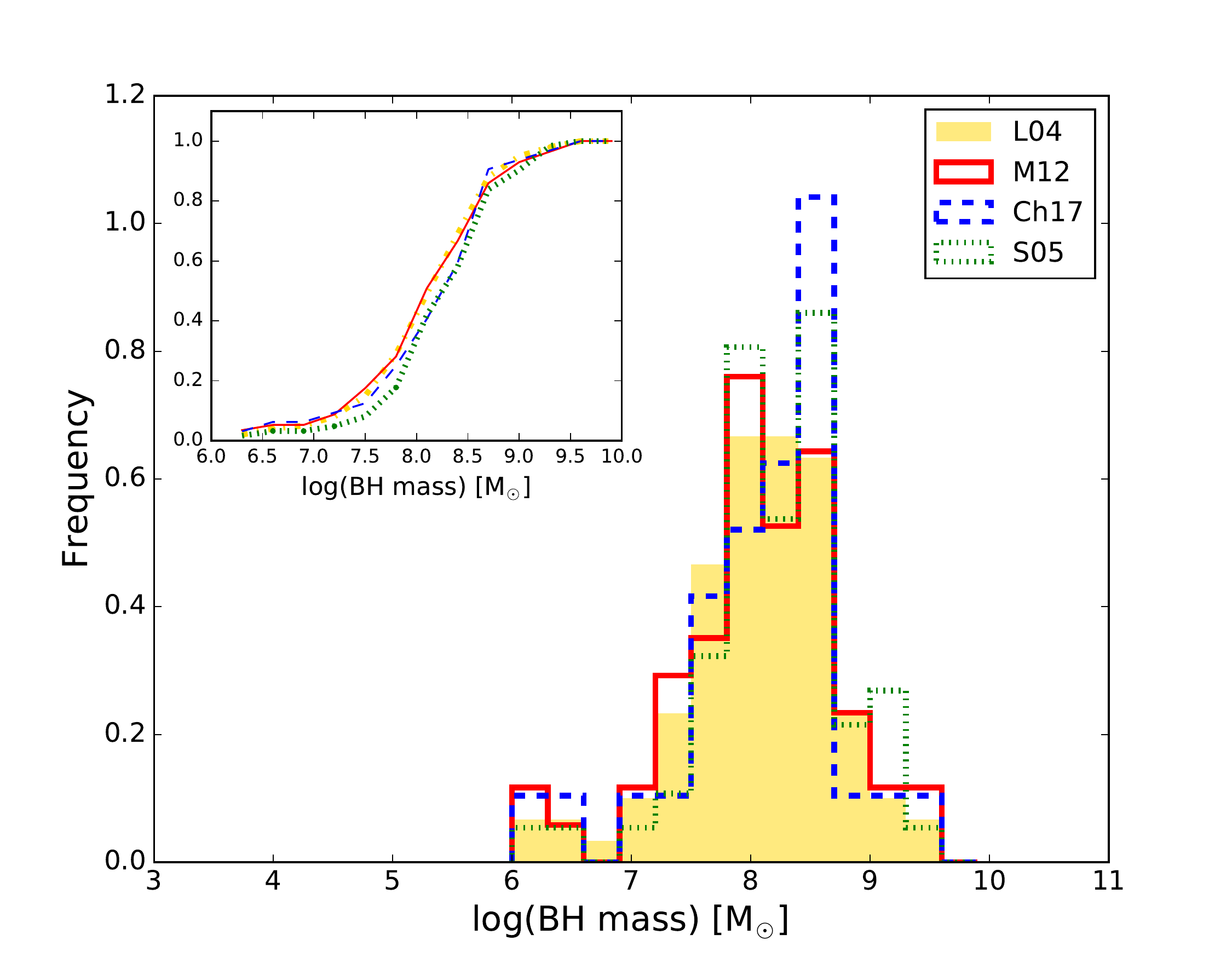}
\caption{Distributions of black hole mass for each MIR and near-IR method using the MEx diagram.  
The inset box represents the cumulative fraction distribution of each MIR and near-IR methods.}  
\label{BHmass}
\end{figure}

In this section we study the black hole mass and accretion properties of AGNs selected using different diagnostic diagrams. 
We have estimated the black hole mass of AGNs from the width of the \oiiil\, following the relation presented in \citet{nelson2000}:

\begin{eqnarray}
\rm log(M_{BH})=(3.7\pm0.7)\times \rm log(\sigma_{\rm \oiii})-(0.5\pm0.1)
\end{eqnarray}

where the M$_{\rm BH}$, the black hole mass is in units of M$_{\odot}$ and $\sigma_{\rm \oiii}$ (calculated as FWHM$_{\rm \oiii}/2.35$) in km/s. 
These authors used the stellar velocity dispersion $\sigma_{\star}$ or
$\sigma_{\rm \oiii}$ in an equivalent way due to their
 assumption that for most AGNs the forbidden line kinematics are dominated by
 virial motion in the host galaxy bulge.

This evidence is based on the results obtained by \citet{nelson96} who found a moderately strong correlation between FWHM$_{\rm \oiii}$ and $\sigma_{\star}$ for the majority of Seyfert galaxies, indicating roughly equal absorption and emission-line widths.


\begin{figure}
  \includegraphics[width=85mm]{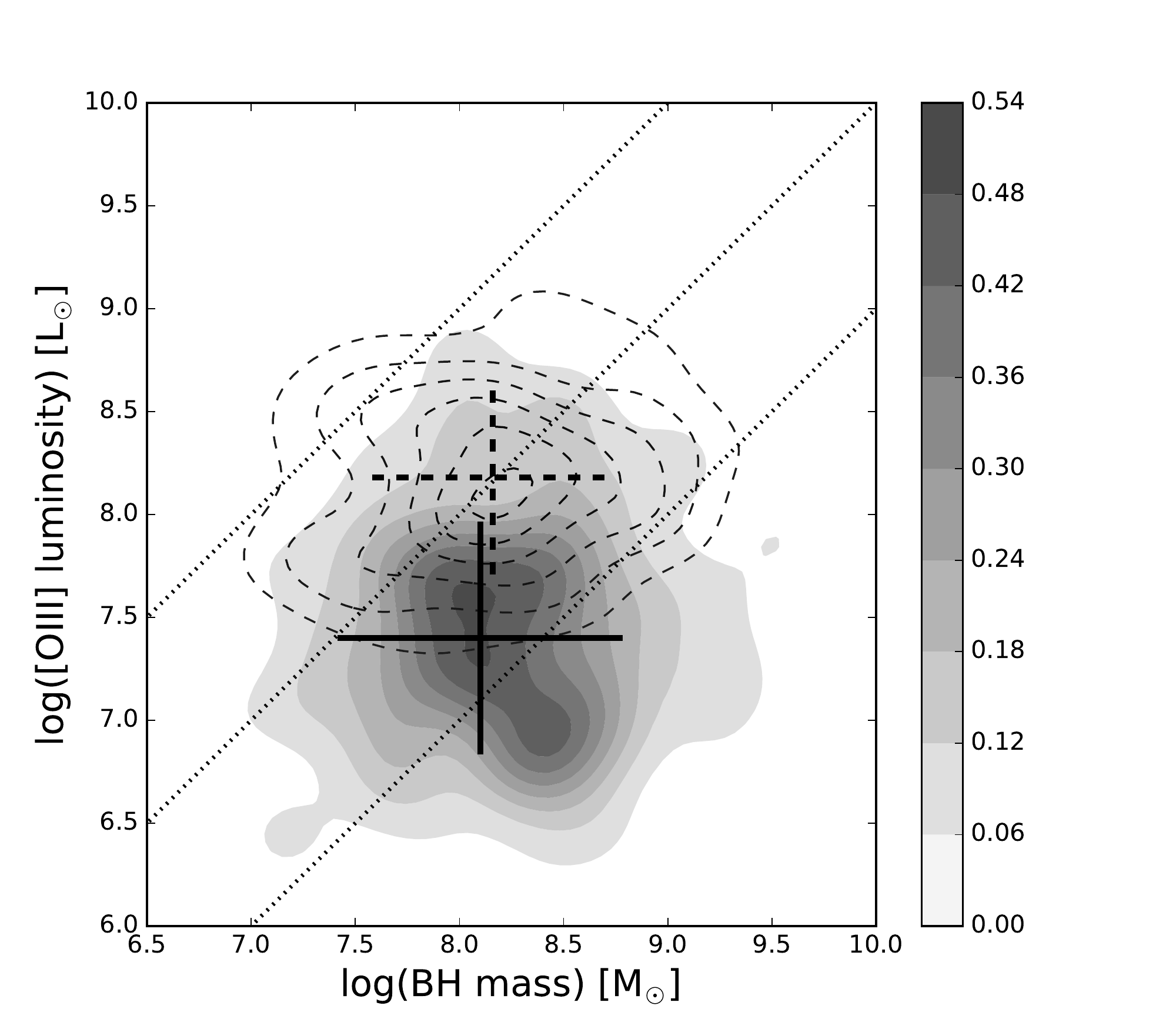}
\caption{OIII $\lambda$5007 luminosity vs black hole mass for AGNs selected according to the MIR and near-IR methods using the MEx line diagnostic diagrams (grey contours). Dashed line contours represent the corresponding values for highly accreting QSOs. Point lines from upper to lower right indicate $\sim$100\%, $\sim$10\%, and $\sim$1\% of the Eddington limit, respectively, assuming bolometric correction 3500 
for OIII luminosity \citep{heckman04,Choi}. Solid and dashed line bars represent the mean and 1$\sigma$ values for our sample of AGNs and for the sample of highly accreting QSOs, respectively. }
\label{cometa}
\end{figure}

Following these premises, in Figure~\ref{BHmass}, we plot the black hole mass distributions for AGNs selected using the MEx diagram according to MIR and near-IR methods of L04, S05, Ch17 and M12.  AGNs selected using the MEx diagram show similar log(BH mass) distributions for all AGNs selected in the MIR and near-IR wavelengths.  We have also included the cumulative fraction distributions for each of MIR and near-IR methods (top left panel), in order
to make more clear the differences between the distributions.

Figure~\ref{cometa} shows grey contours representing the \oiiil\ luminosity vs black hole mass for AGNs selected using the MEx line diagnostic diagrams. 
Dashed line contours are data taken from highly accreting QSOs selected from the SDSS DR7 survey from \citet{negrete18}.
Dotted lines from upper to lower right indicate $\sim$100\%, $\sim$10\%, and $\sim$1\% of the Eddington limit, respectively, assuming bolometric correction of 3500 for L$_{\rm \oiii}$ \protect\citep{heckman04,Choi}. 
These extreme accreting QSOs were selected according to the FWHM of the broad component (BC) of \hb\ (\hb$_{BC}$) and the ratio of equivalent width of \feii\ and \hb$_{BC}$ (specifically with R$_{\rm FeII}>1$, \citealt{negrete18}).
The peak of this distribution coincides with the line indicating sources with $\sim$10\% of the Eddington limit. For the sample of AGNs identified using the MEx diagrams, the peak of the distribution is only above the line correspond for sources with $\sim$1\% of the Eddington limit.
In the figure, our AGN sample according to the MEx diagnostic diagrams have black hole mass distributions similar to those found in highly accreting QSOs at similar redshifts. Although the \oiiil\ luminosity values are on average $\sim$0.7 dex lower than the highly accreting QSO sample.

\subsection{The Morphology}
\label{morpho}

\begin{figure}
  \includegraphics[width=100mm]{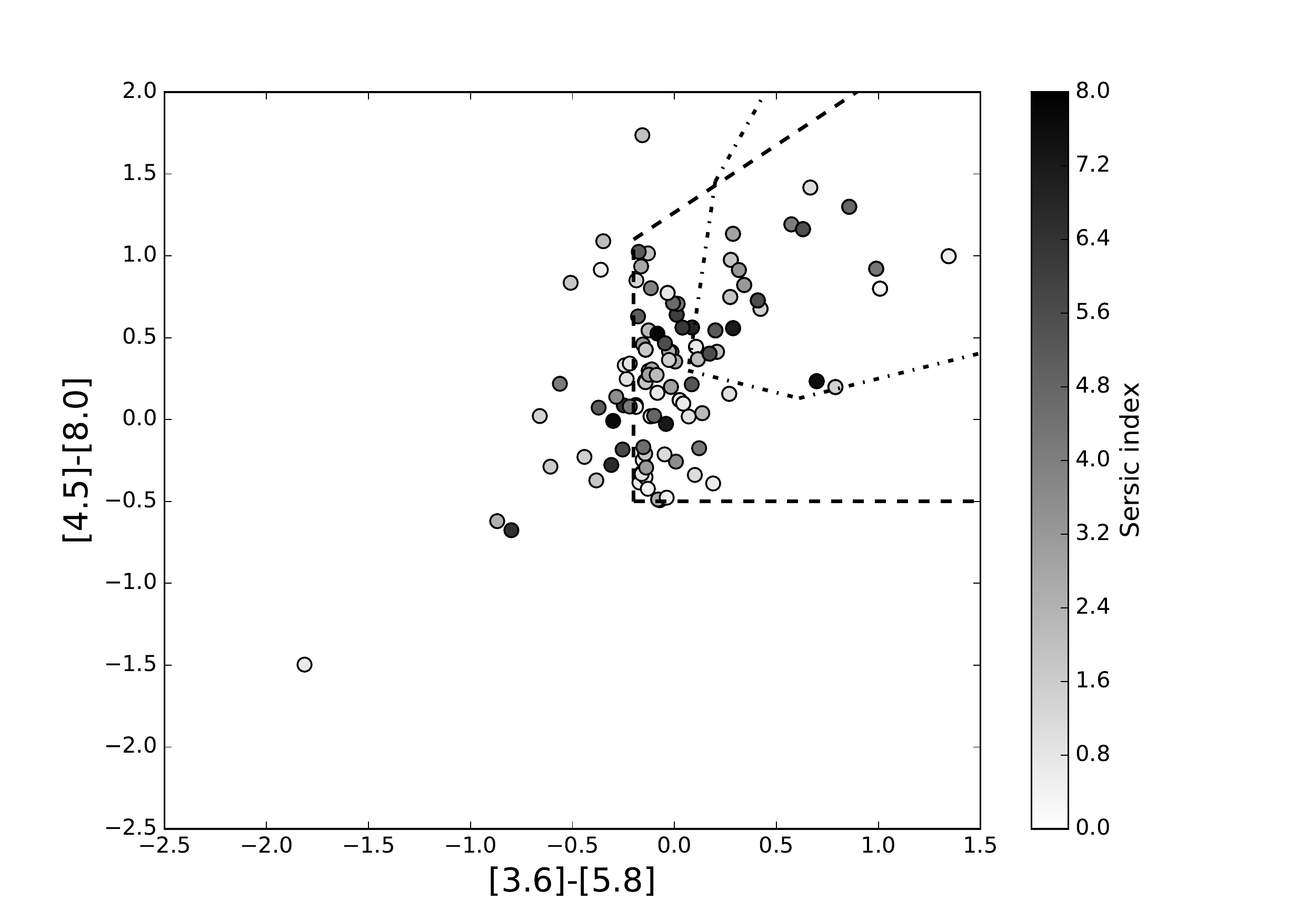}
\caption{MIR colour-colour diagrams for pre-selecting AGNs according the methods presented by L04, S05, Ch17 and M12 using the MEx line diagnostic diagram.  The region marked with dashed and dot-dashed lines show the selection criteria of L04 and Ch17, respectively.  The vertical bar shows the corresponding values for the S\'ersic index.}
\label{nser}
\end{figure}

\begin{table*}
\center
\caption{
Percentage of sources found according to S\'ersic index values, separate in three classes  using the MEx diagram, and IRAC1, IRAC2, XMM1, XMM2, VLA1 and VLA2 samples from \citet{griffith2010}. The G10z sample represent AGNs taken from \citet{griffith2010} within 0.3$\leq z_{\rm ph} \leq$ 0.9. }
\begin{tabular}{ccccccccc}
\hline 
Sample &   MEx & IRAC1 & IRAC2 & XMM1 & XMM2 & VLA1 & VLA2 & G10z\\
\hline
$0.2<n<1.5$             &24.8\%&17\%  & 28\% & 11\% & 16\% & 7\%  & 21\%&21.7\%   \\
 $1.5\leq n \leq 2.5$   &15.1\%& 11\% & 13\% & 4\%  & 11\% & 4\%  & 12\%&11.8\%   \\
 $2.5 < n < 8$          &39.5\%& 41\% & 41\% & 35\% & 44\% & 61\% & 48\%&46.4\%  \\
\hline
\label{table4}
\end{tabular}
\end{table*}
 
An important tool to diagnostic galaxy evolution is the galaxy morphology. In this section we will analyse the different AGN host morphology properties according to parametric methods, such as the S\'ersic index and non-parametric statistics, such as the Gini coefficient and the asymmetry. 

\subsubsection{S\'ersic index}

\citet{gri12} presented a photometric and morphological database using publicly available data obtained with ACS instrument on the $Hubble$ $Space$ $Telescope$, the Advanced Camera for Surveys General Catalog (ACS-GC). These authors calculated morphological parameters such as the S\'ersic index \citep{sersic1968} using an automated fitting method called {\small GALAPAGOS} \citep{galapagos}, which is compound by {\small GALFIT} \citep{peng02} and {\small SExtractor} \citep{bertin} codes.
We cross-correlated our AGN sample with this catalogue 
using a matching radius of 1 arcsec. We found that 97\% of our AGNs selected by the MEx diagram are presented in the catalogue.  In Figure~\ref{nser}, we plot the MIR colour-colour diagram for pre-selected AGNs according to the methods present by L04, S05, Ch17 and M12 for AGNs obtained using the MEx diagram. The region marked with dashed and dot-dashed lines show the selection criteria of L04 and Ch17, respectively. We have included a relative grey scale showing the S\'ersic index values of each source. 

We have restricted the S\'ersic index to be in the range $0.2 < n <  8$ and we have divided the galaxies into three classes, according to \citet{griffith2010}: sources with $0.2 < n < 1.5$ comprised of late-types or spirals, those with $1.5\leq n \leq 2.5$ comprised with galaxies with blended morphologies with bulge+disk components and with $2.5 < n < 8$ generally comprised of ellipticals or early-type galaxies. 
We have not considered extreme S\'ersic model profiles $n=0.2$ and $n=8$ which probably correspond to erroneous fits or systematics.

\citet{griffith2010} investigated the optical morphologies of AGN candidates identified at MIR, X-ray and radio wavelengths. 
They defined 6 samples of AGNs called IRAC1, IRAC2, XMM1, XMM2, VLA1 and VLA2. The IRAC1 sample  consists of the brighter MIR AGN candidates selected
to the 5$\sigma$ depth of the original IRAC Shallow Survey with MIR criteria proposed by \citet{S05}.
The IRAC2 consists of a fainter sample of objects detected in all four IRAC bands down to the full 5$\sigma$ depth of the $Spitzer$
Deep Wide-Field Survey, but not already detected in the IRAC1 sample.
The XMM1 sample consists of sources with soft (0.5-2.0 keV) X-ray fluxes S$_{0.5-2.0}$=5.0$\times$ 10$^{-15}$ erg cm$^{-2}$ s$^{-1}$ and the fainter sample, XMM2, of sources with S$_{0.5-2.0}$ $<$ 5.0$\times$10$^{-15}$ erg cm$^{-2}$ s$^{-1}$.
 The radio VLA1 (Very Large Array 1) consists of brighter sources, with flux densities greater than or equal to 1 mJy (S$_{1.4}\geq$1.0 mJy) and VLA2, of fainter sources, with flux densities within 0.3$\leq$ S$_{1.4}<$ 1.0 mJy.
 We have also constructed a sub-sample of AGNs, that we will call G10z, obtained from the \citet{griffith2010} catalogue using the same redshift cuts (0.3$\leq$ z$_{\rm ph}$ $\leq$ 0.9, estimated using photometric redshifts) as our AGN samples. 
 
 Table~\ref{table4} shows the percentage of sources found according to their S\'ersic index values separate in three classes using the MEx diagram, the bright and faint samples obtained with IRAC, XMM and VLA, and the G10z sample. Comparing our $n$ values with other MIR, X-ray and radio selected AGNs, we find that AGNs selected in the fashion of MEx diagram have similar percentages of
 late-types or spirals (with $0.2<n<1.5$), spiral/late-type galaxies with bulge+disk components (with $1.5\leq n \leq 2.5$) and 
 galaxies with blended morphologies with bulge+disk components with $2.5 < n < 8$ compared to the IRAC1 and IRAC2 selection methods.
 We also find similar percentages between our AGNs for the three classifications according to the S\'ersic index values and the G10z sample with similar redshift distributions.
 Although the sample of weak and bright sources detected in radio wavelengths (VLA1 and VLA2) present a higher percentage of objects with early-type morphologies compared to the sample of AGNs according to the MEx diagram.

\begin{figure}
  \includegraphics[width=80mm]{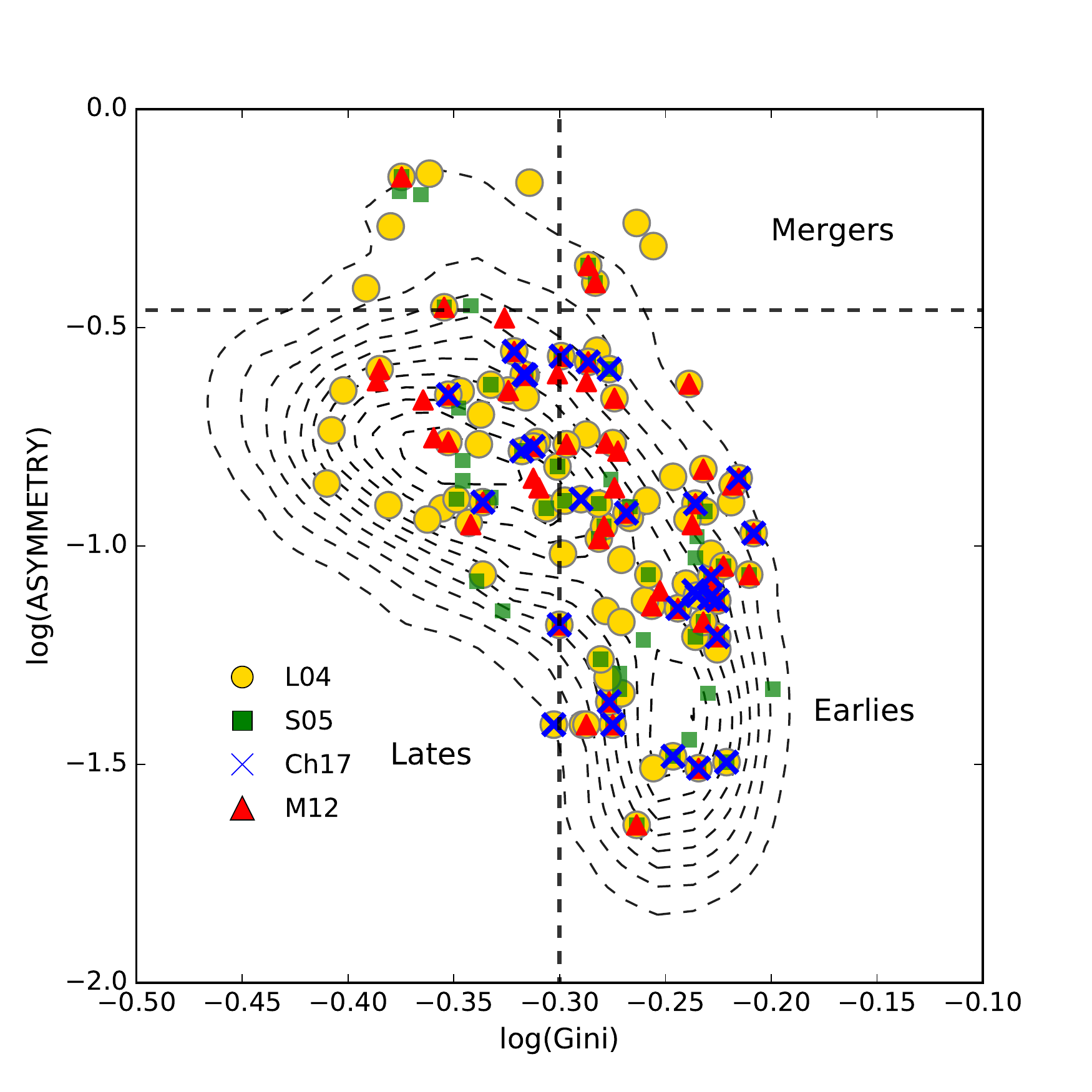}
\caption{log(Asymmetry) vs. log(Gini) coefficient for AGNs selected using the MEx method. Symbols are the same as used in other figures. Dashed lines shows the regions populated by mergers and late and early-types galaxies. Dashed line contours represent galaxies taken from the zCOSMOS-non-{AGN} identified in the \citet{cassata2007} catalogue.}
\label{asy}
\end{figure}

\subsubsection{Gini coefficient and asymmetry index}

Non-parametric approaches to quantitative morphology have been developed over the last years by several authors \citep{abraham1996,conselice2003,lotz2004,cassata2007,tasca2009,lotz2008}.
\citet{conselice2003} presented the CAS parameters: the concentration index, the asymmetry and clumpiness.  The asymmetry parameter A quantifies
the degree to which the light of a galaxy is rotationally
symmetric. The value of A is calculated by rotating a galaxy through 180$^\circ$ and subtracting this rotated galaxy from the original and comparing the absolute value of the residuals of this subtraction to the original galaxy flux \citep{conselice2000}.
Zero asymmetry would correspond to a completely symmetric galaxy, typically of elliptical types with smooth light profiles and Asymmetry=1 would correspond to a totally asymmetric one, such as spiral galaxies, irregular types or galaxies with major merger signatures. 

Later other parameters were added such as the Gini coefficient and the M$_{20}$ parameter \citep{lotz2004,lotz2008}. 
The Gini coefficient, first introduced by \citealt{abraham2003}, is a statistical tool originally used in economics 
for measuring the distribution of wealth within a population, and was found useful in astrophysics for finding the quantitative measurement of inequality
of galaxy light distribution between pixels. Gini=1 would mean that all the light is in one pixel, while Gini=0 would mean everyone/every pixel has an equal share. The Gini coefficient is a good marker of an overall smoothness and as such it can be a good probe of merger process in general. 

The morphological measures used in this work were obtained from the catalogue presented by \citet{cassata2007} which provides information of non-parametric diagnostics
of galaxy structure using the $Hubble$ $Space$ $Telescope$ ACS for 232022 galaxies up to F814W=25 mag.
In Figure~\ref{asy}, we plot the Asymmetry vs the Gini coefficient for AGNs selected using the MEx diagnostic diagram. 
The horizontal dashed lines at log(Asymmetry)$=-$0.46 \citep{conselice2003} shows the dividing line above which objects are expected to be major mergers. While the vertical dotted line at log(Gini)$=-$0.3 \citep{Abraham2007} separate early-type galaxies to the right.  Dashed line contours in the figure represent the morphological values observed for galaxies in the zCOSMOS-non-{\footnotesize AGN} sample.

A visual inspection performed on AGNs using the ACS F814W images shows that spiral galaxies lie preferentially in the low-Gini-low-asymmetry region of the Figure~\ref{asy}, although many spiral galaxies with predominant bulges also occupy the region established for early-types.
This result agrees with some works that show that late-type galaxies occupy a large scattered zone both to the left and to the right (occupied by early-types) of the line log(Gini)$=-$0.3 \citep{Abraham2007, kartaltepe2010}.
Despite this, AGNs denominate the area populated by late-types to that defined by log(Asymmetry)$<-$0.46 and log(Gini) $< -$0.3.
 
Without making distinctions between AGNs pre-selected using the MIR and near-IR methods, we find the following percentages: (early, late, merger) = (67.5, 23.2, 9.3)\%.
For the sample of control galaxies zCOSMOS-non-{\footnotesize AGN} we find (early, late, merger) = (44.3, 49.7, 6)\%.
As it can be seen, the percentage of late-type galaxies in the MEx AGN sample is less than a half when compared to the control sample of galaxies without AGNs. On the other hand, both the percentages of early-types and mergers, are one and a half larger than in the control sample.




\section{Summary and discussions}
\label{sum}
In this paper, we have studied four selection methods in the MIR and near-IR wavelengths: the IRAC colour-colour cuts proposed by \citet{L04} and \citet{S05}, according to a power-law emission \citep{Ch17} and a combination of MIR and near-IR emission and based on the predictions by galaxy and AGN templates \citep{M12}.
We have employed the line diagnostic diagrams which uses
the {\oiii}/{\hb} line ratio and the {\oiiil} line width, the
kinematic-excitation diagram \citep{Zh18} and stellar
mass, the mass-excitation diagram \citep{J11, J14}.

The main results can be summarised as follows:

\begin{itemize} 

\item Within the four MIR and near-IR methods analysed in this work, there are two (L04 and S05) which have high contamination by SF and composite galaxies: between 60-70\% of contamination of SF according to the MEx diagnostic diagram and $\sim$50\% of composite galaxies according to KEx diagram. These methods were also reported by other authors for suffering contamination by SF galaxies \citep{sajina2005,donley12, mendez2013, kirk2013}.

\item The methods with the greatest success in selecting AGNs, according to the diagnostic diagrams, are those presented by M12 and Ch17, with a percentage of success ranging from 48 to 56\%. 
The method proposed by M12 selects a large number of AGNs (63 according to the MEX method) and appears to be reasonably efficient in both the success rate (61\%) and total number of AGN recovered.
These results are in agreement with those found by \citet{M12} who determined with the help of colour tracks for various galaxy type models that the contamination by normal galaxies appears significantly reduced in comparison to L04 and S05 methods.

\item We find that the S05 method selects a considerable percentage of predominantly blue, low-mass and low redshift objects with MIR spectra that follow positive blue slopes ($\alpha$ $\geq$ 0). These are characteristic of stellar-dominated sources at low redshifts, which generally exhibit positive IRAC power-law emission following the Rayleigh-Jeans tail of the black-body spectrum \citep{park2010}. This effect is greater when AGNs are selected according to the KEx method compared to the MEx method, which when using stellar masses allows a better distinction between low-mass galaxies with high star formation from pure AGNs.

\item By analysing the colours in the near-UV and optical ($M_{NUV}-M_r$) and according to the KEx diagram, the method proposed by S05 selects a high fraction of objects identified with SF galaxies with low dust content. 

\item According to the KEx and MEx diagrams, most of the AGNs have $M_{NUV}-M_r$ colour distributions bluer than those found for the green valley galaxies, which is more noticeable in the KEx diagram.
Similar results were found by \citet{hickox2009} in a sample of IRAC selected AGNs using optical colours. These authors found that IRAC selected AGNs are found throughout the galaxy colour–magnitude space, with a few hosts on the red sequence and a predominant population towards blue colours with respect to the green valley.

\item With respect to the KEx and MEx line diagnostic diagrams, from the analysis of the \oiiil\ luminosity and the $M_{NUV}-M_r$ colour, we find that the KEx diagram that uses the width of the \oiiil\ line, probably does not allow a clear separation between the emission of the \oiii\ lines coming from an AGN from that originating in the HII regions of galaxies with high star formation rates.

As noted by \citet{Zh18}, the explanation behind the KEx method would consist in assuming that the width of the \oiiil\ line correlates with the stellar velocity dispersion ($\sigma_{\star}$).
This would be correlated with the mass of the bulge and the stellar mass of the galaxy. Although in the literature there are several works that postulate the correlation between \oiiil\ and $\sigma_{\star}$ \citep{nelson2000, Ho2009, komossa2007}, some work did not find such a close correlation \citep{boroson2003, botte2005, Bennert2018,Sexton2019}. Particularly, \citet{botte2005} found that the width of the \oiiil\ line typically overestimates the stellar velocity dispersion.
Therefore, using the width of the \oiiil\ line would not be a good indicator to separate SF galaxies from pure AGNs.

\item{According to the pre-selection methods in the MIR and near-IR of L04, S05, Ch17 and M12, we find that between 15\% to 40\% of objects have X-ray emission.
Comparing the results obtained in a subsample of sources with 0.3 <z <0.9 and X-ray emission in \citet{civano}, we have found that the AGN samples according to the S05 and M12 methods (also identified according to the MEx diagram) have an excess of between 15 to 21\% of obscured objects (HR$>-$0.2).}

\item AGNs selected using MIR and near-IR methods according to the MEx diagrams have black holes with masses similar to highly accreting QSOs at similar redshifts, but are $\sim$0.7 dex less luminous in \oiiil\ luminosity.


\item The majority of hosts of AGNs identified using MEx diagrams are early type  galaxies. When compared to a suitable control sample with non-AGN galaxies with similar stellar mass and redshift distributions, we find that MEx selected AGNs are 50\% more probably found in early-types, and in galaxies with major merger signatures.

\end{itemize}
\section{Acknowledgements}
We thank the anonymous referee for his/her useful comments and suggestions.

This work was partially supported by the Consejo Nacional de Investigaciones Cient\'{\i}ficas y T\'ecnicas (CONICET) and the Secretar\'ia de Ciencia y Tecnolog\'ia de la Universidad Nacional de C\'ordoba (SeCyT). 
Based on data products from observations made with ESO Telescopes at the La
Silla Paranal Observatory under ESO programme ID 179.A-2005 and on
data products produced by TERAPIX and the Cambridge Astronomy Survey
Unit on behalf of the UltraVISTA consortium.
Based on zCOSMOS observations carried out using the Very Large Telescope at the ESO Paranal 
Observatory under Programme ID: LP175.A-0839 (zCOSMOS).
This research has made use of the VizieR catalogue access tool, CDS,
 Strasbourg, France (DOI : 10.26093/cds/vizier). The original description 
 of the VizieR service was published in 2000, A\&AS 143, 23.




\section{APPENDIX}
In this section we analyse the completeness of the AGN spectroscopic sample as a function of MIR magnitudes and colours.
First, we investigate the dependence of the fraction of sources on spectroscopic redshift and $[8.0]$ $\mu$m magnitude, which is the ultimate selection band in the MIR. We calculate the percentage of objects with $[8.0]$ $\mu$m $<$22.9 (which is the average limiting magnitude according to \citealt{laigle}) and 0.3 $\le z_{\rm phot} \le 0.9$ that have reliable spectroscopic redshifts (sample zCOSMOS-$good$) according to  L04, Ch17, S05 and M12  (43\%, 37\%, 41\% and 35\%, respectively).

\begin{figure*}
  \includegraphics[width=195mm]{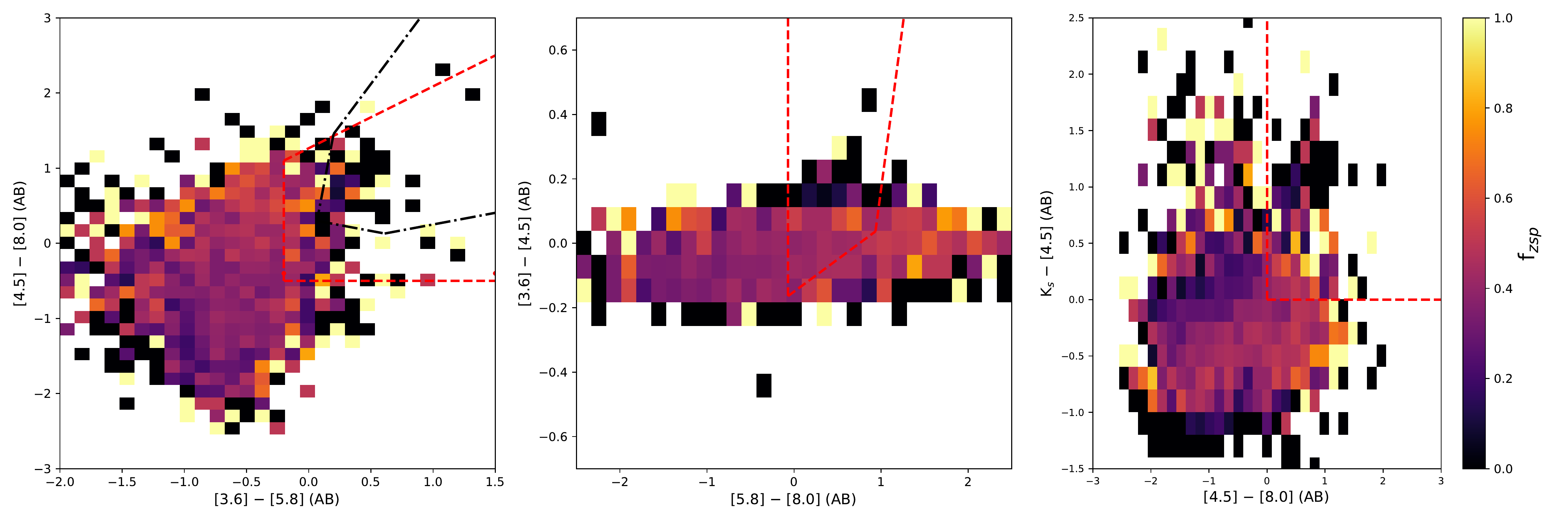}
\caption{MIR and near-IR colour-colour diagrams and the selection of AGNs for different photometric methods of L04 (dashed lines) and Ch17 (dot-dashed lines, left panel), S05 (middle panel) and M12 (right panel).  The vertical bar represents the fraction of objects with good redshift estimates (zCOSMOS-$good$) and the sample of objects within the MIR and near-IR limiting magnitudes and with 0.3$\le$ z$_{\rm phot}$ $\le$ 0.9.}
\label{f_spec}
\end{figure*}

We have also investigated a possible dependence with the colours and the fraction of objects  with spectroscopic redshifts. 
In Figure \ref{f_spec} we plot MIR and near-IR colour-colour diagrams and the selection of AGNs according to the criterion of L04 and Ch17 (left panel), S05  (middle panel) and M12 (right panel). The vertical colour bar represents the fraction of objects with accurate spectroscopic redshifts (zCOSMOS-$good$ sample) and the sample of objects with MIR and near-IR detections and 0.3$\le$ z$_{\rm phot}$ $\le$ 0.9. 

Although, as expected, the fraction of sources with z$_{\rm spec}$ decays for red MIR colours, we find that, on average, the fractions of AGN selected according to the methods of L04, Ch17, S05 and M12 are 0.38, 0.30, 0.36 and 0.33, respectively.
These results show that our AGN samples were selected with no important selection biases among the different methods in the MIR and near-IR.

\end{document}